# Ultra-High Carrier Mobilities in Ferroelectric Domain Wall Corbino Cones at Room Temperature

Conor J. McCluskey[1], Matthew G. Colbear[1], James P. V. McConville[1], Shane J. McCartan[1], Jesi R. Maguire[1], Michele Conroy[2], Kalani Moore[2], Alan Harvey[2], Felix Trier[3,4], Ursel Bangert[2], Alexei Gruverman[5], Manuel Bibes[3], Amit Kumar[1], Raymond G. P. McQuaid[1] and J. Marty Gregg[1*]

[1] School of Mathematics and Physics, Queen's University Belfast, Belfast, BT7 1NN, U.K.
[2] Department of Physics & Bernal Institute, University of Limerick, Limerick, Ireland
[3] Unité Mixte de Physique, CNRS, Thales, Université Paris-Saclay, 91767, Palaiseau, France
[4] Department of Energy Conversion and Storage, Technical University of Denmark, 2800 Kongens Lyngby, Denmark
[5] Department of Physics and Astronomy, University of Nebraska, Lincoln, NE 68588, USA

*email: m.gregg@qub.ac.uk

Recently, electrically conducting heterointerfaces between dissimilar band-insulators (such as lanthanum aluminate and strontium titanate) have attracted considerable research interest. Charge transport has been thoroughly explored and fundamental aspects of conduction firmly established. Perhaps surprisingly, similar insights into conceptually much simpler conducting homointerfaces, such as the domain walls that separate regions of different orientations of electrical polarisation within the same ferroelectric band-insulator, are not nearly so well-developed. Addressing this disparity, we herein report magnetoresistance in approximately conical 180° charged domain walls, which occur in partially switched ferroelectric thin film single crystal lithium niobate. This system is ideal for such measurements: firstly, the conductivity difference between domains and domain walls is extremely and unusually large (a factor of at least $10^{13}$) and hence currents driven through the thin film, between planar top and bottom electrodes, are overwhelmingly channelled along the walls; secondly, when electrical contact is made to the top and bottom of the domain walls and a magnetic field is applied along their cone axes (perpendicular to the thin film surface), then the test geometry mirrors that of a Corbino disc, which is a textbook arrangement for geometric magnetoresistance measurement. Our data imply carriers at the domain walls with extremely high room temperature Hall mobilities of up to ~ 3,700 cm$^2$V$^{-1}$s$^{-1}$. This is an unparalleled value for oxide interfaces (and for bulk oxides too) and is most comparable to mobilities in other systems typically seen at cryogenic, rather than at room, temperature.

# 1. Introduction

The idea that two adjoining electrically insulating materials might generate a new sheet-like conducting state, along the interface between them, is remarkable. However, it is not new. Almost 50 years ago, in a theoretical study, Vul *et al.*[1] considered boundaries between two insulating dielectrics, with opposing senses of spontaneous electrical polarisation; a local accumulation of charge was postulated, of sufficient density that metallic interfacial behaviour was expected. Experimentally, signs of domain wall conduction were later seen (e.g. Schmid and Petermann)[2] and in the nineteen nineties, Aird and Salje[3] even reported superconductivity along specific twin boundaries (in tungsten trioxide). Somehow, these studies failed to prompt immediate follow-up activity from the rest of the scientific community. It was only in 2004, when Ohtomo and Hwang reported distinct electrical conductivity at interfaces between single crystal $SrTiO_3$ substrates and $LaAlO_3$ thin films[4], that the field really exploded. Since then, considerable research has been done and a great deal is now known about the fundamental physics of conduction, along oxide heterointerfaces (those between two different band insulators) in particular.

Homointerfaces (boundaries between distinct regions of the same insulating material) have not been ignored, but the nature of research has been somewhat different. For ferroelectric domain walls in particular, there has been a phase of simply discovering new systems in which enhanced electrical conduction can been seen in both proper ($BiFeO_3$[5,6], $LiNbO_3$[7,8], $BaTiO_3$[9] and $Pb(Zr_xTi_{1-x})O_3$[10]) and improper systems (hexagonal manganites[11], copper chlorine boracite[12] and Ruddlesden-Popper phases[13]). Significant focus has also been given to the manipulation of domain walls: their site-specific controlled injection, removal and movement have been established[12,14–18] and conductance states have been dynamically modified (by tuning their number[19] and inclination angles[20,21]). A primary driver for all of this research has been the possibility of combining 2D electrical conduction with an inherently dynamic nature (domain walls can be created, moved and destroyed by externally applied fields) to realise a new paradigm of reconfigurable "domain wall electronics"[22]: a technology in which domain wall nanocircuitry might be formed, changed and erased on demand. This exciting idea has already motivated the demonstration of a number of different domain wall devices such as bit memories, memristors and transistors[19,23–27].

Fundamental explorations have also been done: evidence has been found for defect aggregation producing inter-band dopant states at ferroelectric domain walls[28], bending of the electronic bands due to polar discontinuities has been postulated[29], and alterations in the electronic band structure have been seen to produce a slight narrowing of the band gap over that established in bulk[5,11,30]. Quantitative aspects of transport behaviour, based on scanning probe microscopy-enabled Hall voltage measurements, have demonstrated relatively high carrier mobilities (of the order of hundreds of $cm^2V^{-1}s^{-1}$)[31,32], in both $ErMnO_3$ and $YbMnO_3$. Rather recently, machine-learning-assisted data analysis has also been employed, untangling the domain wall response from bulk and interface effects[33].

Taken as a whole, however, insight into the fundamental physics of transport in homointerfacial ferroelectric domain walls is distinctly underdeveloped and has been somewhat in the shadow of the technology-driven work alluded to above. This is particularly evident when progress is compared to the state-of-the-art in heterointerfacial oxide boundary research.

Here, we seek to redress this imbalance by reporting magnetoresistance measurements on charged conducting domain walls, created by partial switching, in ion-sliced single crystal lithium niobate ($LiNbO_3$) thin films (500nm thick). The domain walls are approximately conical in morphology and we show that this allows magnetotransport to be interpreted in terms of the classical Corbino disc geometry[34–37], which is distinct from the magnetoresistance seen previously at conducting domain walls in $BiFeO_3$[38,39]. Our geometric magnetoresistance measurements imply an exceptionally high room temperature domain wall Hall mobility, of the order of thousands of $cm^2V^{-1}s^{-1}$. This is dramatically larger than reported in other oxide semiconductors[40–42] and heterointerfaces[4,43], where the highest room temperature mobilities observed are in the region of tens to hundreds of $cm^2V^{-1}s^{-1}$. Transport comparisons are perhaps better made with intrinsic elemental semiconductors (Si = 1350 $cm^2V^{-1}s^{-1}$, Ge =3600 $cm^2V^{-1}s^{-1}$)[44] or two-dimensional electron gases at low temperatures. We note, however, that hints of extremely high carrier mobilities in $LiNbO_3$ have been seen previously, in photocurrent experiments[45,46]. Our observations, while dramatic, are therefore not completely without precedent. Nevertheless, they do suggest extraordinary transport behaviour in these domain walls. In addition to mobility, we also infer the carrier density to be a small fraction ($10^{-5}$) of that required for screening the polar discontinuity at the wall and this indicates, as has been seen in other domain wall systems[31,32], that the overwhelming majority of assumed screening charges do not participate in conduction.

**2. $LiNbO_3$ (LNO) domain wall morphology.**

Our studies were performed on commercially-obtained 500nm thick ion-cut single crystal thin films, with 150nm thin film gold-chromium bottom electrodes. The uniaxial polarisation in these films is perpendicular to their surfaces (z-cut). As-received, the LNO was in a monodomain state. However, partial switching of polarisation, under an applied electric field (using an additional top electrode), caused needle-like domains to grow. After traversing the interelectrode gap, needle domains grow radially outwards, maintaining a rather strong inclination with respect to the polar axis. This results in a truncated cone morphology across the interelectrode gap. Fig. 1a shows the circular base sections (dark contrast) of a number of domain wall truncated cones, evident on the top surface of a LNO film, along with associated electrical conduction footprints (Fig. 1b). Note that current maps are somewhat smeared, but this does not indicate interelectrode conduction through domains themselves. It instead reflects the finite size of the conducting atomic force microscopy (AFM) tip (radius ∼ 20-40nm), maintaining electrical contact with domain walls, even when the centre of the tip has moved past them, as well as the possibility of low resistance pathways that involve traversing a short section of LNO at the surface, to reach strongly conducting domain wall source-drain conduits. PFM and cAFM images in literature (for example in [19]) show similar smearing effects, but scans using new sharp tips systematically reveal that conduction arises at the domain wall (see supplementary information S10). While these specific domain patterns were written using a raster-scanned AFM tip as the top electrode (a sharp-point field source), we observed a similar microstructure, even when poling with a liquid metal Indium-Gallium-Tin eutectic mesoscale planar top electrode (Fig. 1c). Conical structures were confirmed in cross section, by the appearance of domain contrast in approximately isosceles triangular forms (Fig.1d). The microstructure, which we have schematically represented in Fig. 1e is consistent with various other domain investigations across both bulk and thin film LNO samples[19,20,24,47].

**3. LNO domain walls as Corbino cones.**

Current travelling through a solid is deflected by a magnetic field, due to the magnetic component of the Lorentz force. The consequences of this deflection depend on the sample geometry. In a Hall bar, for example, carriers are deflected towards insulating sample edges, where they accumulate, to develop an electric field which builds to counteract the magnetic field-induced deflection. Ideally, in equilibrium, carriers return to their straight-line path along the external electric field direction, and a Hall potential is maintained on the transverse sample boundaries. If no insulating sample boundaries exist, then no Hall field can develop, the magnetic deflection of the current persists, and carriers deviate from their shortest path between source and sink electrodes. This increase in carrier path length leads to a magnetic field-dependent increase in the sample resistance, termed geometric magnetoresistance[35]. The magnitude of the magnetic force on the moving carrier is governed by its drift velocity in the electric field. Therefore, the size of the magnetic deflection, and resulting resistance change, can be used to infer carrier mobility[48–50]. This approach is commonly used to characterise the mobility active in short channel device geometries, where the sample is much wider than it is long[51–55] ($l/w \ll 1$, where $l$ is the length between current carrying electrodes and $w$ is the channel width). In this geometry, known as the "short Hall bar", many carriers, which are deflected by a Lorentz force, reach the sink electrode before encountering a sample boundary. Thus, the full Hall potential does not develop, and carrier deflection persists. Intermediate geometries have also been studied[56–58]: in rectangular samples, where $l/w < 3$, some Hall potential is maintained on the sample boundaries, but it is insufficient to fully counteract the magnetically induced deflection, and a (reduced) MR persists.

While the ideal Hall bar ($l/w = \infty$) is not realisable, the ideal geometric magnetoresistor is. It is the well-known Corbino disc[35,56] geometry (Fig. 2c): an annular sample sandwiched between two concentric electrodes. This geometry is the limiting case of the short Hall bar ($l/w = 0$). It completely mitigates any carrier build-up, as the only boundaries present are at the electrodes. No Hall field develops and a full geometric magnetoresistance is manifested; Hall fields are effectively "shorted".

The equivalence of the conical conducting domain walls, connecting top and bottom electrodes, and the Corbino disc geometry can be loosely inferred by geometric projection (onto the plane parallel to the LNO sample surface): the conical conducting LNO domain wall is then the Corbino annulus and the domain wall intersections with the top and bottom electrodes form the concentric outer and inner contacts (black and yellow rings in Fig. 1e), illustrated schematically in Fig. 2 b, c. It is important, however, to explicitly derive the form of the magnetoresistance response for the cone.

Typically, geometric magnetoresistance varies quadratically with applied magnetic field[35]. This conclusion can be drawn from the analytic solution for the full current density, in the presence of magnetic and electric fields (**B** and **E**) in a Drude-type model of conduction (supplementary information notes S1-S3). A Hall field can be included, but only if its form is known. For the conical domain wall geometry, with no Hall correction considered, this analytic solution suggests that current pathways will deviate from the domain wall itself and into bulk LNO. Given its highly insulating nature, this is unphysical and so we conclude that a small Hall potential must develop across the width of the domain wall, creating a Hall field perpendicular to the domain wall surface, to confine the carriers inside the conducting conduit.

To accurately assess the full current density in the conical geometry, including an estimation of any Hall field, we use an iterative summation procedure. Here, we consider the action of the magnetic part of the Lorentz force on carriers as a series of successive, diminishing current components, which are corrections to the conventional current drift in an electric field, directed along the major axes in the system. Each successive correction is a higher order of the principal deflection $\mu B$, with the sum converging if $\mu B < 1$. This allows us to account for Hall fields by applying geometry-specific boundary conditions: current components directed off the cone are cancelled by a Hall component, while current remaining on the cone is allowed, and this component forms the basis of the next iteration. This approach is taken from that presented in reference [35]. Its applicability is discussed further in the supplementary information (S4, S5), but the essence and implied form of the geometric magnetoresistance response is summarised below.

The electric and magnetic fields acting on the current-carrying domain wall truncated cones can be represented in cylindrical coordinates as

$$E_0 = \begin{pmatrix} E_r \\ E_\varphi \\ E_z \end{pmatrix} = E_0 \begin{pmatrix} -\cos\theta \\ 0 \\ -\sin\theta \end{pmatrix} \quad (1)$$

$$B = \begin{pmatrix} B_r \\ B_\varphi \\ B_z \end{pmatrix} = B_0 \begin{pmatrix} 0 \\ 0 \\ -1 \end{pmatrix} \tag{2}$$

where $r, \varphi$ and $z$ represent the radial, azimuthal and $z$ axis bases of cylindrical coordinates, and θ is the inclination angle of the domain wall, defined as the acute angle between the domain wall surface normal and the z axis. $E_0$ and $B_0$ are the magnitudes of the electric and magnetic fields respectively.

Walking through the first few iterations of this cycle reveals the key behaviour. Initially, current moves along the electric field direction which defines the cone surface:

$$\boldsymbol{j_0} = \sigma_0 \boldsymbol{E_0} \tag{3}$$

The first deflection is then along $\boldsymbol{E_0} \times \boldsymbol{B}$, which is azimuthal in this case:

$$\boldsymbol{j_1} = \sigma_0 \boldsymbol{E_1} = \sigma_0(\boldsymbol{v_{d1}} \times \boldsymbol{B}) = \sigma_0 \mu (\boldsymbol{E_0} \times \boldsymbol{B}) = \mu(\boldsymbol{j_0} \times \boldsymbol{B}) = \sigma_0 E_0 \begin{pmatrix} 0 \\ -\mu B_0 \cos\theta \\ 0 \end{pmatrix} \tag{4}$$

Here, $E_1$ is an "effective electric field", which would be that needed to generate the carrier deflection that is actually caused by the magnetic component of the Lorentz field, and $v_{d1}$ is the drift velocity due to this effective field. $j_1$ is azimuthal and remains on the cone, so no Hall correction is required yet.

The next deflection suffered by the carriers is radial:

$$\boldsymbol{j_2} = \mu(\boldsymbol{j_1} \times \boldsymbol{B}) = \sigma_0 E_0 \begin{pmatrix} \mu^2 B_0^2 \cos\theta \\ 0 \\ 0 \end{pmatrix} \tag{5}$$

and clearly does not remain on the cone surface (Fig. 2b). Given that carriers cannot propagate into the highly insulating bulk LNO (as discussed above), this component illustrates that a Hall potential must form across the width of the domain wall itself. It is in this detail that the current behaviour in the conical system differs from that in the true planar Corbino disc: while we still have an increase in path length due to the azimuthal component of equation (4), and thus expect a geometric magnetoresistance, we also expect a small Hall potential, justifying the use of the iterative investigation and approach.

$j_2$ gives information on the form of the magnetoresistance and Hall potential. Projecting $j_2$ onto the external electric field direction, we find a current component opposing the initial current direction (shown in purple in Fig. 2b):

$$j_{2,E} = \frac{j_2 \cdot E_0}{|E_0|^2} E_0 = \sigma_0 E_0 \begin{pmatrix} \mu^2 B_0^2 \cos^3\theta \\ 0 \\ \mu^2 B_0^2 \sin\theta \cos^2\theta \end{pmatrix} = -\mu^2 B_0^2 \cos^2\theta \, \sigma_0 E_0 \tag{6}$$

This is the first approximation of the magnetoresistance. Only this allowed current component is then used in the next series of iterations. The remaining component of $j_2$ ($j_{2,Norm}$, red arrow in Fig. 2b) is orientated along the cone surface normal and will first develop and then be countered by a Hall potential, as outlined above. Following the procedure developed in equations (3) to (6) through a few more iterations (shown in the supplementary information), a full current density in the electric field direction is as follows:

$$j_{B,E} = (1 - \mu^2 B_0^2 \cos^2\theta + \mu^4 B_0^4 \cos^4\theta \dots) \sigma_0 E \tag{7}$$

This approximates to a full solution of:

$$j_{B,E} \approx \frac{\sigma_0 E}{1 + \mu^2 B_0^2 \cos^2\theta} \tag{8}$$

We can then deduce the magnetoresistance, for a z-oriented magnetic field, as:

$$MR = \frac{R_B - R_0}{R_0} = \frac{j_0 - j_{B,E}}{j_{B,E}} = \mu^2 B_0^2 \cos^2\theta \tag{9}$$

Despite the Hall potential that must form across the width of the domain wall in the conical geometry, we still develop the same kind of quadratic variation of magnetoresistance with magnetic field, as is seen in the Corbino Disc. The correction factor of $\cos^2\theta$ is the same as one might expect from considering only the component of current in the cone which is influenced by magnetic field, i.e. by ignoring the $z$-component of current and projecting the cone onto the plane perpendicular to the cone axes (visualised in Fig. 2b,c).

Quantitative interpretations of geometric magnetoresistance are restricted by a low field limit, which dictates that $\mu B$ is small (see supplementary information note S2). Therefore, the expected resistance change might only be of the order of 1% or less, depending on the carrier mobility. To confidently isolate such a small signal, a periodic, linearly varying magnetic field profile was applied to the sample over several days, with current measurements taken continuously. The periodicity of the applied field (triangular waveform) allowed identification of the magnetoresistance signal in frequency space (Fig. 3b) as well as an averaging of the current response over many magnetic field cycles. The current response (averaged over 20 cycles) is plotted in Fig. 3a along with the magnetic field profile used. A decrease in current with increasing magnetic field strength is clearly visible. Importantly, current varies with the magnitude of the applied magnetic field and is independent of its sign. Moreover, its form is parabolic (following equation (9)). Results from two separate experiments with differing frequencies of applied magnetic field cycles are shown in Fig. 3b. Clearly, the current response exhibits a principal component at the second harmonic frequency, as expected. Fig. 3c shows the parabolic form of the magnetoresistance explicitly with a fitted $B^2$ coefficient of $(6.1\pm0.5) \times 10^{-3}$. This was the largest magnetoresistance that was exhibited across all samples studied (presumably with the lowest contact resistance, as a large magnetically inactive in-series contact resistance will act to drown the MR signal). Fig. 3d expresses the same information logarithmically, allowing direct extraction of the squared exponent in B.

To determine the carrier mobility, we apply equation (9), assuming an angle of inclination $\theta$ of approximately 78°. This gives a corrected room temperature mobility of $\mu \approx 3700 cm^2 V^{-1} s^{-1}$, an unusually high value for oxide semiconductors at room temperature. The estimated inclination angle is consistent with microstructural observations shown in Fig. 1d, and with geometrical arguments associated with the domain diameter needed for significant source-drain currents to develop [19].

The magnetoresistance measurement was repeated on a separate sample, with different applied voltages (Fig 4a) used to induce different current densities in the domain walls. No observable change in the magnetoresistance is seen and so it appears that the carrier mobility is insensitive to current density variations (at least at these applied voltage levels which do not cause further ferroelectric switching).

The inferred mobility measurement allows for an estimation of carrier density, provided the current density is known. Other in-house studies and published works[19] have observed huge currents through LNO domain walls, of the order of 0.1A for an array of domains similar to that shown in Fig. 1. Taking cylindrical domain arrays with radius of order $r$ = $10^{-7}$m and estimating a top electrode size of the order $A_E \sim 1 \times 10^{-9} m^2$, with a driving field $E$ = $10^7$ Vm$^{-1}$ (typical values used when driving ~0.1A through our two-terminal multi-domain walls structures), we can estimate the active carrier density ($n_{active}$) as

$$n_{active} = \frac{I}{A_E \frac{2\pi r d}{\pi r^2} E \mu e} \approx \frac{10^{13}}{d} m^{-3} = 10^9 cm^{-2} \qquad (10)$$

where $\mu$ is the carrier mobility, $e$ the electronic charge and $d$ the domain wall thickness, in meters. Note that this is a low-end estimate, as any effects from parasitic contact resistance are not included in equation 10. Contact resistance will act to reduce the measured MR signal and hence also reduce the inferred mobility. An estimate of the carrier density in two dimensions implied by full screening of the polar discontinuity at the wall is:

$$n_{screen} = \frac{2P.S}{eS} \approx 10^{14} cm^{-2} \tag{11}$$

where $P$ is the spontaneous polarisation of LNO (~0.7Cm$^{-2}$) and again a domain wall inclination angle of 78° between the polarisation and the domain wall surface normal is used. The ratio of $n_{active}$ to $n_{screen}$ (~10$^{-5}$) indicates that a relatively small fraction of the screening charge, assumed to be present, actually contributes to conduction, commensurate with previously obtained Hall effect measurements on domain walls in ErMnO$_3$[32].

**4. Implications of the magnetoresistance observations.**

We observe a quadratic fractional increase in domain wall resistance with increasing magnetic field strength. While the quadratic response is characteristic of geometric magnetoresistance (as discussed and derived above), there are additional magnetotransport phenomena which could also be at play. In particular, the so-called "physical magnetoresistance" is known to often have a quadratic dependence on magnetic field strength[35,56,59] and could exist in our "Corbino cones", in addition to the geometric effect. The challenge is then to disentangle magnetoresistance contributions, due to the absolute magnitude of mean carrier mobility (geometric), from those due to a spread in mobility values (physical). Prior studies [35,49,56,60] conclude that, when both effects are simultaneously present, geometric magnetoresistance dominates the overall behaviour.

Furthermore, previous domain wall magnetotransport studies found either no magnetoresistance[38], or an effect with a magnitude inconsistent with physical magnetoresistance[39], suggesting that significant spread in carrier mobility is not a general property of domain wall transport. In the minority of cases, in other materials systems, where the carrier mobility spread is greater than the mean mobility, a non-saturating linear magnetoresistance is expected[61] and should persist into the low-field regime. We see no hint of linearity in our measurements. Other domain wall studies, involving much larger fields[38], have seen no linear effects either and this is commensurate with the view that carrier mobilities are much greater than the spread in carrier mobilities in conducting domain wall systems generally. Extraordinary MR can occur in specialised metal-semiconductor hybrid systems. Neither the nature of our system, nor the magnitude of the observed MR, is consistent with an extraordinary MR effect.

Fig 4b shows the effect of changing the angle between the applied magnetic field and the polar axis of the LNO films (the z axis in Fig 2b). The magnitude of the magnetoresistance measured in this instance was generally lower than shown in figure 3. We attribute this to a larger contact resistance and inferior wire-bond. Nevertheless, data clearly show that there is no significant variation in the magnetoresistance with magnetic field orientation. This surprising observation demands explanation. We have therefore derived an expression for the geometric magnetoresistance (again using an iterative approach), for magnetic fields applied parallel to the x-y plane ($MR_{xy}$). The full treatment is given in the supplementary information (section S9). It is more complicated than that associated with the field oriented parallel to the polar axis, as Lorentz forces cause diametrically opposite strips of carrier accumulation and denudation to develop (supplementary information, figure S9b). This builds a Hall potential, such that the conical domain walls act as two "half-cone" short Hall bars, connected electrically in parallel. Short Hall bars, with intermediate length-to-width ratios, allow geometric magnetoresistance to persist, albeit reduced by a prefector dependent on aspect ratio. Using our derived form of the $MR_{xy}$ (supplementary information S9), along with an analysis from Lippmann and Kuhrt (to estimate the aspect ratio prefactor)[58], we find the resultant magnetoresistance (for the x-y plane oriented field) to be:

$$MR_{xy} = \frac{0.25 \, \mu^2 {B_0}^2 sin^2\theta}{2} \qquad (12)$$

Using this expression and assuming the geometric magnetoresistance, most accurately represented in figure 3, to be constant with respect to magnetic field orientation (as implied by figure 4b) yields an alternative estimate of room temperature carrier mobility of ~2,000cm$^2$V$^{-1}$s$^{-1}$; this is lower than that given by equation 9, but by less than a factor of two (reasonable given the coarse nature of some of the assumptions made in deriving equation 12).

Ultra-high carrier mobility, accompanied by a low carrier density, should allow insight into the possible charge transport mechanisms at play. Defect-related electron hopping seems unlikely (low carrier mobility is observed when hopping transport is dominant). Furthermore, defect aggregation at the domain walls is slow in LNO: defects seem largely insensitive to electric fields at room temperature (and so would not migrate to the charged wall to provide screening); this is evidenced by an extremely low bulk dark conductivity[62] ($10^{-16} - 10^{-18}$ Ω$^{-1}$cm$^{-1}$). In any case, charged defect migration to domain walls has previously been blamed for a reduction, rather than increase, in conductivity[63].

Fundamental alteration in the electronic band-structure at the LNO domain wall may be possible. A reduction of the band gap, as has been seen in BFO[30], would clearly lead to an increase in the electron population in the conduction band at room temperature. The ultra-high carrier mobilities observed may also indicate a strong curvature developing at the bottom of the conduction band (which is not present in bulk LNO[64]), consistent with low effective carrier masses. Notable transport behaviour has been seen at surfaces in conventional bulk semiconductors before, particularly at heterointerfaces. Like LNO, perovskite $KTaO_3$ (KTO) displays a similarly innocuous electronic band structure in bulk[65,66]. However, more detailed band structure measurements[67] on the (111) face revealed a 2DEG with the potential for exotic transport properties. Superconductivity, for example, has been observed along heterointerfaces of (111) planes in KTO and EuO or LAO at low temperatures[68].

Even metallic conduction can be considered. Current-voltage measurements as a function of temperature (supplementary section S8 and refs [63,69]) make this speculative, but the two probe measurement geometry suffers from series contact resistances, which may obscure any observation of metallicity. We discuss the theoretical possibility of a metal insulator transition in the supplementary information (S8), and note that further investigation of domain wall transport by 4-probe methods is required to give greater insight.

## 5. Summary and Outlook.

In summary, we have performed novel geometric magnetoresistance measurements on conical conducting domain walls in lithium niobate. These measurements represent the first detailed magnetotransport study in this 2D system, and suggest a uniquely large room temperature mobility. Potential conduction mechanisms are discussed, but further and similar measurements at cryogenic temperatures are needed to fully elucidate the transport in these systems.

## 6. Experimental Section

**Sample preparation**

The sample consists of a 500nm thick z-cut single crystal of lithium niobate, commercially available from NanoLN and prepared by the ion-slicing method[70]. It includes a gold bottom electrode and a $SiO_2$-LNO insulating substrate. For most of the current measurements, including the magnetoresistance measurements, silver thin film electrodes (approximately 20nm thick) were deposited onto the surface of the LNO film through hardmask grids in 100x100µm$^2$ squares via thermal evaporation; the bottom electrode (already present in the as-received samples) was contacted by coating the side with conducting silver electrodag. 25µm diameter aluminium wire was wire-bonded to the silver top electrode by the application of ultrasonic pulses, providing robust electrical connections for the experiment. The as-grown wafer is monodomain with its polarisation pointing away from the bottom electrode. Conducting domain walls were bias written, with the sample connected in series to a 100kΩ current-limiting resistor. This both protects the electrode from high current-density pulses during the switching process and produces a domain structure of sufficient resistance (~ 5MΩ), such that we can keep the current level relatively low (around 1µA at 5V) throughout the MR experiment. A series of 'writing' voltage pulses was applied to the series circuit from 0V to 50V, and back down to 0V in steps of 1V, using a Keithley 237 source-measure unit. After each pulse, a 'reading' pulse of 5V was applied, to check the conducting state of the sample. It was found that a voltage pulse of roughly 26V was enough to produce obvious domain wall conducting pathways through the sample.

**Piezoresponse force microscopy (PFM) / Conducting atomic force microscopy (cAFM)**
PFM and topography measurements were carried out simultaneously with an MFP-3D infinity AFM system from Asylum research. Conducting silver paste contacted the bottom electrode of the sample and was grounded, while an AC voltage is applied to a standard Pt coated Si AFM tip. The frequency of the AC voltage was tuned to match the resonant frequency of the tip-sample system (roughly 330kHz) and the amplitude of the applied signal varies between 1-4V. For cAFM measurements, a steady DC bias of the order of -5V is applied to the LNO bottom electrode while the tip, in contact with the LNO surface, is held at ground. Current is then measured as a function of position on the sample surface.

**Magnetoresistance measurements**
LNO domain walls

The sample was placed between the electromagnet pole pieces, where the magnetic induction may be varied in strength, up to a magnitude of 1 Tesla, and monitored via an external Hall probe. In the standard measurement, current was driven through the conducting domain walls via the application of a constant 5V bias to the surface silver electrode, using a Keithley 237 source-measure unit, and measured as a function of magnetic field. We allowed the current through the sample to stabilise for 2 hours to a noise level of roughly ± 10nA before applying the magnetic field profile shown in Fig. 3a. Given the low field limit mentioned above, we expect the magnitude of the magnetoresistance signal to be small, comparable to the noise level in the system. For this reason, we applied 20 cycles of the B field profile shown in Fig. 3a, which consists of 20 linearly spaced field steps between +/-1T. Each field is applied for 600s, bringing the total time for the experiment to ~ 67h. Repetition of the current measurements and magnetic field cycles allowed us to improve on the ±10nA noise to produce the uncertainties in Figure 3 and 4. The error bars reflect the standard error on measurements of current across the multiple field cycles. Given the long timeline for the experiment, long-scale current variations were also apparent. The current response of each magnetic field cycle was normalised with respect to the base current (I(B=0)) for that cycle. Current measurements were made continuously at each field step, and the current value at any magnetic field is the average of 800 datapoints taken during that step. The average of corresponding current measurements across 20 magnetic field cycles produces the current data in Fig. 3. Various current densities were investigated by applying larger voltages to a similar sample (Fig. 4a). The field angle was investigated by rotating the sample mounting stub within the field.

Control experiment

A control test was also carried out. The magnetoresistance experiment was repeated, with current shorted through the bottom electrode of the same LNO sample used in the main experiment (Fig 3), effectively excluding the LNO film and domain walls from the circuit. No magnetoresistance was seen, eliminating the possibility that observations were due to spurious signal or experimental artefact. These control results are shown in the supplementary information (section S6).

**Current component calculation**

The calculation of current components under electric and magnetic fields was carried out following a method described by Popovic[35]. The rationale is as follows: carriers are initially assumed to move under the influence of the electric field only, giving a current density. Once the carriers drift, they experience the magnetic part of the Lorentz force, giving a second current density component, perpendicular to the first. This component is then deflected, and so on, and is calculated by $j_{n+1} = \mu(j_n \times B)$. Successive current "deflections" are summed, giving a final current density which is accurate to higher powers of $\mu B$, depending on the degree of iteration. We show in the supplementary information how this method provides a solution which converges towards the analytic solution for current density in the Corbino disc.

To solve for the current in the Corbino cone, we further consider insulating boundary conditions: any current component that arises which is orientated towards an insulating boundary is cancelled by an equal and opposite "Hall" component. Using this method, we arrive at the equation for MR in the conical conducting domain walls.

**Transmission electron microscopy**

Cross sectional transmission electron microscopy (TEM) samples were prepared by ion milling, using a dual beam focused ion beam (FIB) microscope. When thinned to electron transparency, the sample was mounted onto an Omniprobe® copper lift-out grid. The scanning TEM (STEM) imaging was performed using a Thermo-Fisher Scientific double tilt TEM holder in the Thermo-Fisher Scientific FEI double aberration corrected Titan Themis Z, located at the University of Limerick, operated at 300kV.

**Data availability**

The data that support the findings of this study are available from the corresponding authors upon reasonable request.

**Acknowledgements**

The authors are grateful for funding support from the Engineering and Physical Sciences Research Council (EPSRC) through grant EP/P02453X/1 and through studentship funding, the UKRI Future Leaders Fellowship programme (MR/T043172/1) and the US-Ireland R&D Partnership Programme (USI 120).

**Author contributions**

The magnetoresistance experimental design, data collection and data analyses were carried out by C.J.McC and M.G.C. Electrical characterisation with temperature was performed by C.J.McC. The details of the MR model were derived by C. J. McC and J.P.V.McC. PFM and cAFM were carried out by J.R.M. and J.P.V.McC. STEM measurements, elucidating the domain wall microstructure, were carried out by M.C., K.M., A.H., U.B. and J.P.V.McC. The experimental idea was conceived by J.P.V.McC and J.M.G. J.M.G supervised the project along with R.G.P.McQ and A.K. M.B. contributed expertise in magnetoresistance measurement and interpretation. All authors contributed to the discussion and interpretation of results, and all were involved in the manuscript preparation.

**Competing interests**

The authors declare no competing interests.

**Figures**

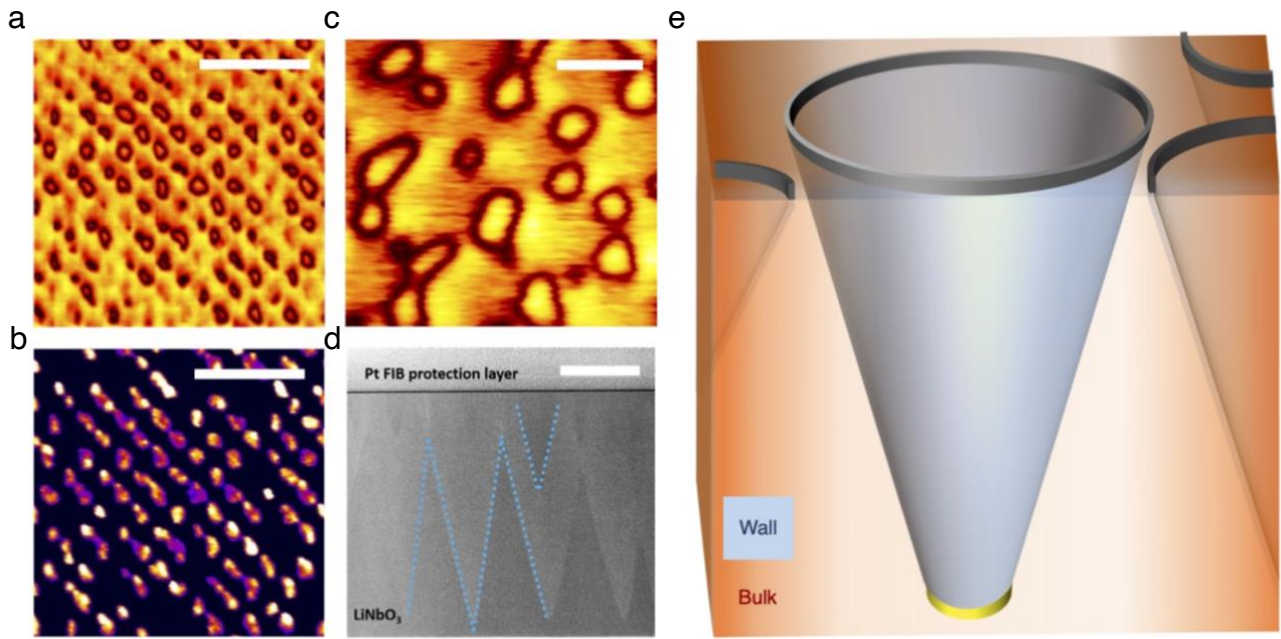

**Figure 1 | Conducting, conical domain walls in LiNbO$_3$. a,** Piezoresponse force microscopy (PFM) amplitude and **b,** conducting-atomic force microscopy (cAFM) maps of domains obtained on the top surface of 500nm thick partially switched lithium niobate. Domains were created using a rastered AFM tip as a top electrode. The scale bar represents 700nm. **c,** PFM amplitude from the top surface of a similar LiNbO$_3$ (LNO) thin film sample after partial switching using a mesoscale liquid top electrode. The scale bar represents 150nm. **d,** Cross-sectional high angle annular dark-field scanning transmission electron microscopy (HAADF STEM) image of the domains in LNO. The overlaid lines highlight the inclination of the walls. The scale bar represents 200nm. **e,** A schematic of the conducting domain wall cones in LNO, as inferred from the top surface PFM in **a** and **c** and the cross sectional HAADF STEM in **d**. The dark grey and yellow rings highlight the locus of the electrical contacts made with the top and bottom electrodes respectively.

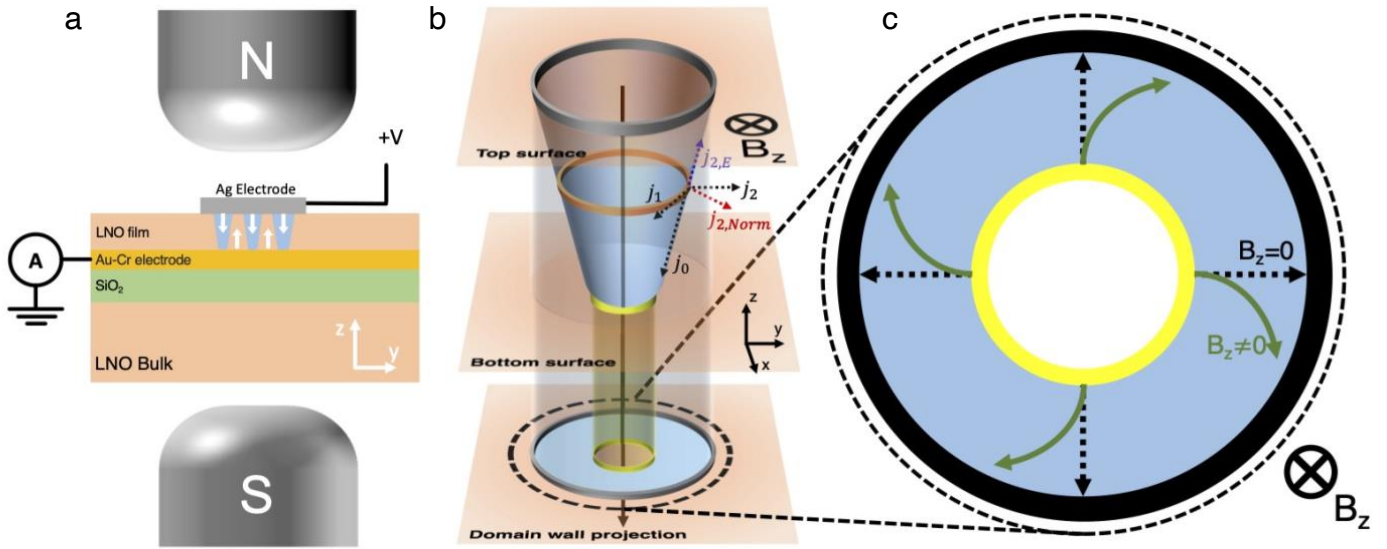

**Figure 2 | Domain walls as Corbino discs. a,** Experimental setup for the geometric magnetoresistance measurement. **b,** Schematic depicting conducting domain wall in LNO, along with its 2D projection, which takes the form of the Corbino disc. The black arrow shows the direction of the magnetic field used in the geometric MR measurement depicted in **a**, and the various dotted arrows show the current components found through an iterative description of the current density. **c,** The typical Corbino disc geometry, employed for geometric magnetoresistance measurements. It consists of a sample under investigation (blue annulus) and concentric inner and outer electrodes (black and yellow rings). Electronic motion in the presence and absence of a perpendicular magnetic field is indicated by full black and dashed green arrows respectively.

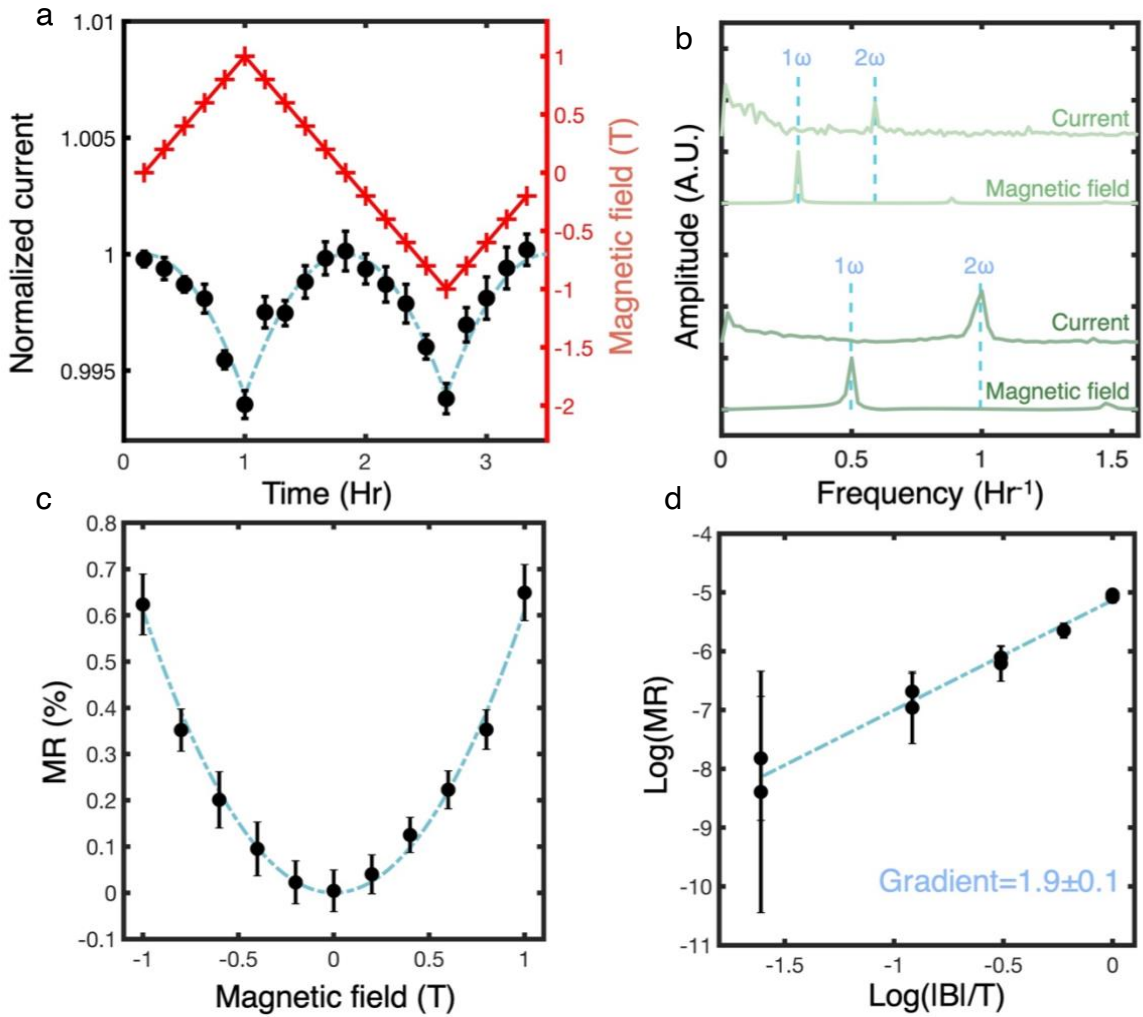

**Figure 3 | Geometric magnetoresistance measurement (MR) at conducting LNO domain walls. a,** The MR response. The red "plus" motifs show the applied external magnetic field and the black circles show the normalised current response, averaged over 20 cycles. The blue dashed lines are parabolic fits (equation (9)). Note that all fits are weighted by the uncertainty on each datapoint. **b,** Discrete fast Fourier transforms of the measured raw current and magnetic field profiles, shown for two datasets. **c,** A plot of the magnetic field vs the MR. The blue dashed line is a fitted parabola with $B^2$ coefficient $(6.1 \pm 0.5) \times 10^{-3}$. **d,** A plot of log(B) vs log(MR), with a linear fit of coefficient $1.9 \pm 0.1$, illustrating the quadratic nature of the B-MR relationship, as expected from our geometric MR treatment.

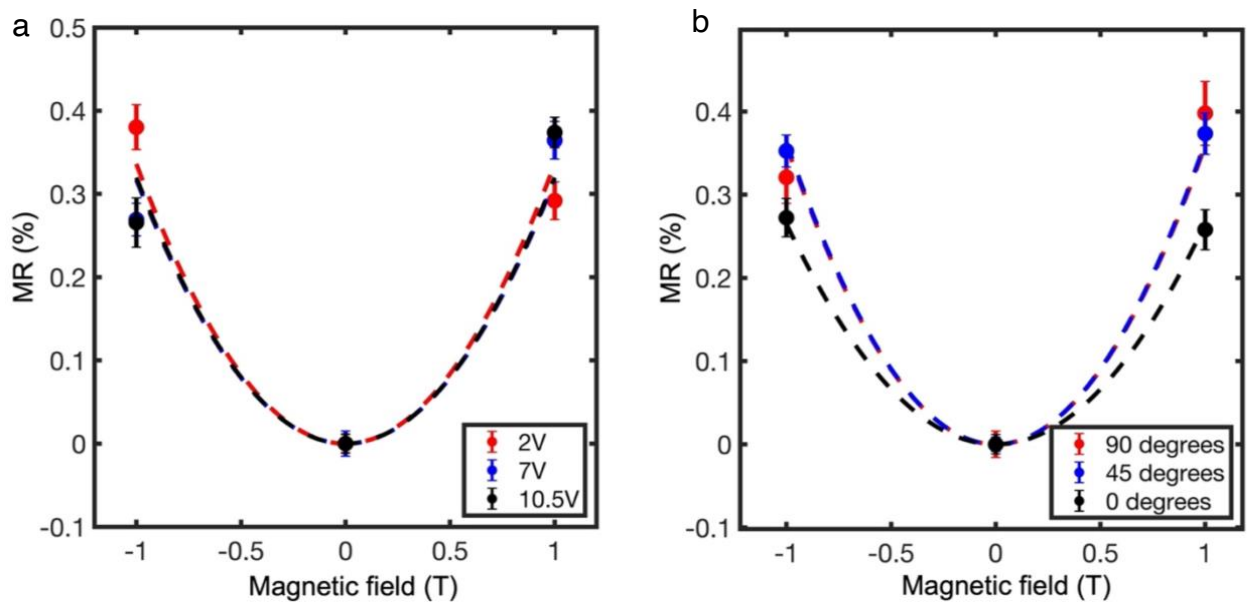

**Figure 4 | Further MR measurements a,** Magnetic field vs MR for LiNbO$_3$ domain wall conduction, as a function of various applied voltages (and therefore as a function of current density). **b,** MR vs magnetic field for various orientations of the magnetic field. At 0 degrees, the field is in the same orientation as illustrated in Fig. 2, and at 90 degrees, the field is in the x-y- plane.

# Ultra-High Carrier Mobilities in Ferroelectric Domain Wall Corbino Cones at Room Temperature

**Supplementary information**

Conor J. McCluskey, Matthew G. Colbear, James P. V. McConville, Shane J. McCartan, Jesi R. Maguire, Michele Conroy, Kalani Moore, Alan Harvey, Felix Trier, Ursel Bangert, Alexei Gruverman, Manuel Bibes, Amit Kumar, Raymond G. P. McQuaid and J. Marty Gregg

**Section S1: Derivation of current density in a material in the presence of E and B fields**

The following derivation follows Popovic[1]. In relatively low fields, we approximate the continuous acceleration and scattering of the charge carrier (of charge $-q$ and effective mass $m^*$) as a smooth drift, with drift velocity $\boldsymbol{v_d} = -\mu \boldsymbol{E_0}$, where $\mu$ is the carrier mobility and $\boldsymbol{E_0}$ is the external electric field. Writing the Lorentz force due to external electric and magnetic fields ($\boldsymbol{E_0}$ and $\boldsymbol{B}$) as an effective electric field $\boldsymbol{E}_{eff}$, we state:

$$-q\boldsymbol{E}_{eff} = -q\boldsymbol{E_0} - q(\boldsymbol{v_d} \times \boldsymbol{B}) \qquad \text{S 1}$$

Note here that the magnetisation in non-magnetic LiNbO$_3$ (LNO) is zero, such that an external magnetic field $\boldsymbol{H}$ and magnetic induction $\boldsymbol{B}$ are proportional. Writing the Lorentz force as an equivalent electrical force allows us to readily represent the resultant current density in the material, in terms of external fields. By multiplying by $-\mu n$, we can write:

$$\boldsymbol{j_B} = \boldsymbol{j_0} - \mu(\boldsymbol{j_B} \times \boldsymbol{B}) \qquad \text{S 2}$$

where $\mu$ is the carrier mobility, and $n$ the carrier density. $\boldsymbol{j_B}$ and $\boldsymbol{j_0}$ are the current densities in the presence and absence of magnetic field, respectively. The dot product of $\boldsymbol{j_B}$ with $\boldsymbol{B}$ is

$$\boldsymbol{j_B} \cdot \boldsymbol{B} = \boldsymbol{j_0} \cdot \boldsymbol{B}. \qquad \text{S 3}$$

while the cross product of $\boldsymbol{j_B}$ with $\boldsymbol{B}$ is:

$$\boldsymbol{j_B} \times \boldsymbol{B} = \boldsymbol{j_0} \times \boldsymbol{B} - \mu[(\boldsymbol{j_B} \times \boldsymbol{B}) \times \boldsymbol{B}] \qquad \text{S 4}$$

Using the vector identity:

$$(\boldsymbol{A} \times \boldsymbol{B}) \times \boldsymbol{C} = \boldsymbol{B}(\boldsymbol{A} \cdot \boldsymbol{C}) - \boldsymbol{A}(\boldsymbol{B} \cdot \boldsymbol{C}) \qquad \text{S 5}$$

and equation S 4, we can write:

$$\boldsymbol{j_B} \times \boldsymbol{B} = \boldsymbol{j_0} \times \boldsymbol{B} - \mu[\boldsymbol{B}(\boldsymbol{j_B} \cdot \boldsymbol{B}) - \boldsymbol{j_B}(\boldsymbol{B} \cdot \boldsymbol{B})] \qquad \text{S 6}$$

After replacing $\boldsymbol{j_B} \cdot \boldsymbol{B}$ using equation S 3, S 5 becomes

$$\boldsymbol{j_B} \times \boldsymbol{B} = [\boldsymbol{j_0} \times \boldsymbol{B}] - \mu[\boldsymbol{B}(\boldsymbol{j_0} \cdot \boldsymbol{B}) - \boldsymbol{j_B}(\boldsymbol{B} \cdot \boldsymbol{B})] \qquad \text{S 7}$$

Substituting S 7 into S 2 we arrive at:

$$\boldsymbol{j_B} = \frac{\boldsymbol{j_0} - \mu(\boldsymbol{j_0} \times \boldsymbol{B}) + \mu^2 \boldsymbol{B}(\boldsymbol{j_0} \cdot \boldsymbol{B})}{1 + \mu^2 B^2} \qquad \text{S 8}$$

In terms of the external fields:

$$\boldsymbol{j_B} = \frac{\sigma_0 \boldsymbol{E_0} + \sigma_0 \mu (\boldsymbol{E_0} \times \boldsymbol{B}) + \mu^2 \sigma_0 \boldsymbol{B}(\boldsymbol{E_0} \cdot \boldsymbol{B})}{1 + \mu^2 \boldsymbol{B}^2} \qquad \text{S 9}$$

if we change $\mu$ such that it carries the sign of carriers. Here, $\sigma_0$ is the conductivity of the material when $B = 0$. This is the result quoted in the main text. Note that, if we include a Hall field contribution, we can replace $\boldsymbol{E_0}$, in equation S 9, with $\boldsymbol{E_0} + \boldsymbol{E_H}$, where $\boldsymbol{E_H}$ is the Hall field in the system.

**Section S2: The low-field condition**

To recover the typical magnetoresistance (MR) response[1]:

$$MR = \mu^2 B^2 \qquad \text{S 10}$$

a smooth drift approximation and low field condition are needed. We now quantify this condition by realising that the applicability of the smooth drift approximation is related to the fraction of a full cyclotron orbit that a carrier can complete between any two scattering events. If the mean free collision time $\langle\tau\rangle$ is much smaller than the time for a cyclotron orbit ($T_C$), the drift is correctly identified as smooth, and regular collisions represent a smooth frictional type force. Physically, as the particle begins its cyclotron orbit under the crossed $\boldsymbol{B}$ and $\boldsymbol{E}$ fields, it collides with the lattice, loses its energy and starts again before completing a significant portion of the orbit. This process repeats, and the resultant is approximately a straight-line path, inclined with respect to the external electric field, with a magnetoresistance response that follows S 10. If several cyclotron orbits are completed between two collisions, the effect of the magnetic field on the carrier is clearly more severe, and our smooth drift breaks down[1]. In the high field case, the MR behaviour therefore departs from S 10. Given the cyclotron orbital period:

$$T_C = \frac{2\pi m^*}{qB} \qquad \text{S 11}$$

and the definition of carrier mobility $\mu$ in terms of mean free transit time between scattering collisions $\langle\tau\rangle$

$$\mu = \frac{q\langle\tau\rangle}{m^*} \qquad \text{S 12}$$

the "low field condition", defined by $\langle\tau\rangle \ll T_c$, becomes:

$$\mu B \ll 2\pi \qquad \text{S 13}$$

**Section S3: MR and Hall electric field in the Corbino disc**

The main consideration in the extension from the Corbino disc to a cone geometry is that the $\boldsymbol{E_0} \cdot \boldsymbol{B}$ term can be zero in the disc (for the typical case of $\boldsymbol{E} \vdash \boldsymbol{B}$ ), whereas for current confined to the cone, this is necessarily not true. The cone is differentiated from a disc by a z height, through which current must pass to traverse along the walls, with z also being the axis parallel to the magnetic field. Therefore, there is a component of current along the magnetic field direction. The analytic result of equation S 9 is general, representing the current density in a material for arbitrarily aligned $\boldsymbol{E}$ and $\boldsymbol{B}$ fields and in the absence of Hall potential build up. Here, we consider the effect of this general solution in disc-like and conical geometries explicitly, showing the similarities and differences.

Beginning with the simpler case of the disc, set up such that $\boldsymbol{E} \perp \boldsymbol{B}$, we use cylindrical coordinates to reflect the symmetry of the system. Here, the driving electric field is radial and the magnetic field is along z:

$$\boldsymbol{E_0} = \begin{pmatrix} E_r \\ E_\varphi \\ E_z \end{pmatrix} = E_0 \begin{pmatrix} 1 \\ 0 \\ 0 \end{pmatrix} \qquad \text{S14}$$

$$\boldsymbol{B} = \begin{pmatrix} B_r \\ B_\varphi \\ B_z \end{pmatrix} = B_0 \begin{pmatrix} 0 \\ 0 \\ 1 \end{pmatrix} \qquad \text{S15}$$

where $r$, $\varphi$ and $z$ denote the usual radial, azimuthal and vertical unit vectors of cylindrical coordinates. $E_0$ and $B_0$ are electric and magnetic field amplitudes. Using the analytic result from S 9, we find a current density of

$$\boldsymbol{j}_{disc} = \frac{\sigma_0 E_0}{1 + \mu^2 B_0^2} \begin{pmatrix} 1 \\ -\mu B_0 \\ 0 \end{pmatrix} \qquad \text{S16}$$

This is the classic result obtained for the magnetoresistance in a Corbino disc. The current density along $\hat{r}$ (the external electric field direction, giving the device current) is reduced by a factor $\frac{1}{1+\mu^2 B_0^2}$, allowing for the appearance of an azimuthal component along $\hat{\varphi}$ which is dependent on the magnetic field strength. As mentioned, no Hall potential builds in the case of the Corbino disc with a surface-normal magnetic field. This means that the solution in S16 is exact.

**Section S4: An iterative approach for calculating current density in the Corbino disc**

It is useful to analyse this same problem using an iterative step like approach[1], as applied by Popovic, because this is the method which can be used to analyse situations beyond the Corbino disc, where Hall potentials can form. In this iterative method, an initial estimate for current is given as the usual drift due to the electric field. Then, Lorentz deflection of each current component that appears is calculated, and progressively smaller corrections to the current density are made. Given that the full solution is accurate for the disc, we here use the iterative solution and compare it to the analytic solution S16.

Starting with $\boldsymbol{j_0}$ as the ordinary radial current component due to the electric field:

$$\boldsymbol{j_0} = \sigma_0 \boldsymbol{E_0} = \sigma_0 E_0 \begin{pmatrix} 1 \\ 0 \\ 0 \end{pmatrix} \qquad \text{S 17}$$

the Lorentz deflection of this current component produces $\boldsymbol{j_1}$, the azimuthal component, which is orientated along $\boldsymbol{j_0} \times \boldsymbol{B}$. Here we write $\boldsymbol{j_1}$ in terms of an effective electric field $\boldsymbol{E_1}$, which can be interpreted as the electric field that would produce the same force on the particle as the magnetic part of the Lorentz force acting on the carrier:

$$q\boldsymbol{E}_{eff} = q(\boldsymbol{v}_d \times \boldsymbol{B}) \qquad \text{S 18}$$

Then we can rewrite $\boldsymbol{j_1}$ in terms of the externally applied fields:

$$\boldsymbol{j_1} = \sigma_0 \boldsymbol{E_1} = \sigma_0 (\boldsymbol{v}_{d0} \times \boldsymbol{B}) = \sigma_0 \mu\, (\boldsymbol{E_0} \times \boldsymbol{B}) = \mu (\boldsymbol{j_0} \times \boldsymbol{B}) \qquad \text{S 19}$$

Here, $\boldsymbol{v}_{d0}$ is the drift velocity of the original current component $\boldsymbol{j_0}$, upon which the Lorentz force acts.

Evaluating $\boldsymbol{j_1}$ gives:

$$\boldsymbol{j_1} = \mu(\boldsymbol{j_0} \times \boldsymbol{B}) = \sigma_0 E_0 \begin{pmatrix} 0 \\ -\mu B_0 \\ 0 \end{pmatrix} \qquad \text{S 20}$$

Subsequent terms can be determined in the same way as in S 19. The next term is the Lorentz deflection of the azimuthal component $j_1$, and produces a third component, $j_2$, opposing the original motion of the carrier:

$$j_2 = \mu(j_1 \times B) = \sigma_0 E_0 \begin{pmatrix} -\mu^2 B_0^2 \\ 0 \\ 0 \end{pmatrix} \quad \text{S 21}$$

This leads to a reduction in current density along $E_0$, and gives the first approximation of the geometric magnetoresistance. The next term $j_3$ adds a correction to the azimuthal current:

$$j_3 = \mu(j_2 \times B) = \sigma_0 E_0 \begin{pmatrix} 0 \\ \mu^3 B_0^3 \\ 0 \end{pmatrix} \quad \text{S 22}$$

The process can be continued to develop further higher order corrections to the current.

If we continue the cycle, we obtain a solution of the form

$$j_{disc} = \sigma_0 E_0 \begin{pmatrix} (1 - \mu^2 B_0^2 + \mu^4 B_0^4 - \mu^6 B_0^6 \ldots) \\ -\mu B_0 (1 - \mu^2 B_0^2 + \mu^4 B_0^4 - \ldots) \\ 0 \end{pmatrix} \quad \text{S 23}$$

Where we can make use of the expansion around $x = 0$:

$$\frac{1}{(1+x)} = 1 - x + x^2 - x^3 \ldots \quad \text{S 24}$$

with $x = \mu^2 B_0^2$ to retrieve the full solution:

$$j_{disc} = \frac{\sigma_0 E_0}{(1 + \mu^2 B_0^2)} \begin{pmatrix} 1 \\ -\mu B_0 \\ 0 \end{pmatrix} \quad \text{S 25}$$

As an illustration to help visualise the approach, the first 3 components of the iterative solution are shown below in fig. S1.

The advantage of this iterative method is in detecting unexpected Hall field components. Any appearing current component that takes the carrier off the conducting surface and onto a sample boundary or insulating part of the sample can be accounted for immediately by including a mitigating Hall potential. That component and its subsequent Lorentz deflection then do not appear in calculations of further components. The full solution assumes the carrier is free to move in any direction, inaccurately predicting the current density when a Hall potential needs to be considered. A correct Hall field, if it is known, can be included in the analytic solution to give the correct current density (by simply replacing $E_0$ in S 9 with $(E_0 + E_{Hall})$). However, as of yet, we have not found a way to accurately assess the Hall field from the analytic solution alone. While the Corbino disc requires no Hall field correction, the case of the cone does, as is discussed below.

It is worth noting that while the methodology of the iterative solution appears to suggest that subsequent Lorentz deflections are occurring in time on a particle, this is not the physical picture. We are calculating, to higher order accuracy, the components of the effect of the magnetic field on a particle, directed along the major axes in the system.

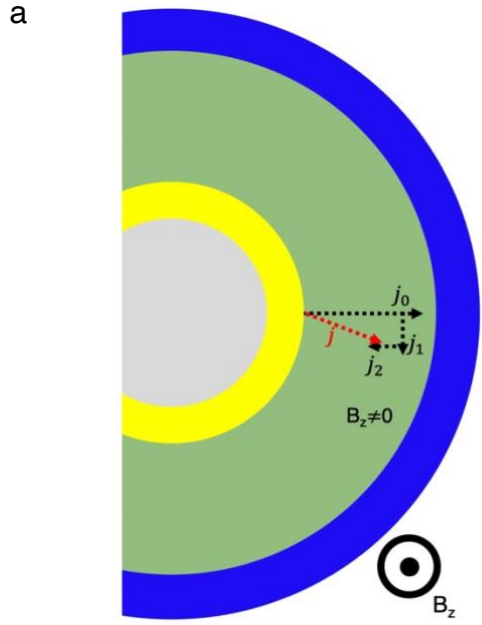

**Figure S1l Iterative solution in the Corbino disc. a,** Vectors of current density produced in the approximate description given above, for a current carrying Corbino disc in the presence of a magnetic field perpendicular to the disc surface.

### Section S5: MR and Hall electric field in the cone

The main difference between the disc and cone is that, in order to drive current along the inclined cone, we now must have a *z* component of the applied electric field. Given that the *z* direction is the applied magnetic field direction, we now must have a component of current along the magnetic field direction, and the $\boldsymbol{E} \cdot \boldsymbol{B}$ term in S 9 no longer disappears. As we show below, if we assume that carriers, subjected to a conical electric field profile, are free to move in any direction (by using the analytic solution with no Hall field included), we find that the resultant current density will take carriers off the original conical surface.

Driving current along a general cone, we have electric and magnetic fields:

$$\boldsymbol{E_0} = \begin{pmatrix} E_r \\ E_\varphi \\ E_z \end{pmatrix} = E_0 \begin{pmatrix} -\cos\theta \\ 0 \\ -\sin\theta \end{pmatrix} \qquad \text{S 26}$$

$$\boldsymbol{B} = \begin{pmatrix} B_r \\ B_\varphi \\ B_z \end{pmatrix} = B_0 \begin{pmatrix} 0 \\ 0 \\ -1 \end{pmatrix} \qquad \text{S 27}$$

The inclination angle is $\alpha$, defined as

$$\tan\alpha = \left|\frac{E_z}{E_r}\right| = \frac{\sin\theta}{\cos\theta} = \tan\theta \qquad \text{S 28}$$

Applying the analytic solution:

$$\boldsymbol{j_{cone}} = \frac{\sigma_0 E_0}{1+\mu^2 B_0^2}\begin{pmatrix}-\cos\theta\\-\mu B_0\cos\theta\\0\end{pmatrix} + \sigma_0 E_0 \begin{pmatrix}0\\0\\-\sin\theta\end{pmatrix} \quad \text{S 30}$$

Immediately, we notice that the radial component of the current density is reduced by a factor of $\frac{1}{1+\mu^2 B_0^2}$, making way for an azimuthal component which is dependent on the magnetic field. In this way the disc and cone are the same. The deviation between disc and cone can be seen when the $z$ component of the current density, which remains unaffected by the magnetic field, is included. Its inclusion pushes carriers off the cone surface, as the ratio of $j_r$ to $j_z$ has changed. This effectively changes the inclination of the cone that would be swept out by carriers if they were free to move in any locus.

$$\tan\alpha = \left|\frac{E_z}{E_r}\right| = \frac{(1+\mu^2 B_0^2)\sin\theta}{\cos\theta} = (1+\mu^2 B_0^2)\tan\theta \quad \text{S 31}$$

Of course, the conducting pathway is actually confined to the conical domain wall which, as far as we know, doesn't change shape during the magnetoresistance measurement. Given that lithium niobate is highly insulating, carrier deviation from the conducting cone will instead be counteracted by the formation of a Hall potential, which builds up across the conical domain wall width.

Therefore, we use an iterative solution to solve for the current density, allowing us to assess both Hall and magnetoresistive components which arise. $\boldsymbol{j_0}$, the component of conventional drift due to the electric field, now acts along the inclined wall of the cone:

$$\boldsymbol{j_0} = \sigma_0 E_0 \begin{pmatrix}-\cos\theta\\0\\-\sin\theta\end{pmatrix} \quad \text{S 32}$$

The Lorentz deflection of $\boldsymbol{j_0}$ produces $\boldsymbol{j_1}$, acting in the direction of $\boldsymbol{E_0}\times\boldsymbol{B}$. Thus, $\boldsymbol{j_1}$ acts in the azimuthal direction, which is tangential to the cone surface. $\boldsymbol{j_1}$ is illustrated in Fig. S2a-b.

$$\boldsymbol{j_1} = \sigma_0 E_0 \begin{pmatrix}0\\-\mu B_0\cos\theta\\0\end{pmatrix} \quad \text{S 33}$$

This initial deflection would keep the carrier on the cone, as at all points $\boldsymbol{E_0}\times\boldsymbol{B}$ locally follows the surface of the cone, and this case is identical to the disc. Thus, we need not consider a Hall potential yet. The Lorentz deflection of $\boldsymbol{j_1}$ produces a force along $\boldsymbol{j_1}\times\boldsymbol{B}$, creating a $\boldsymbol{j_2}$. In the cone scenario, $\boldsymbol{j_1}\times\boldsymbol{B}$ is orientated **radially outwards** from the conical axis (note that this is not perpendicular to the cone surface), as shown in Fig. S2a-b. This component $\boldsymbol{j_2}$ is clearly not embedded in the cone, and so would result in a Hall potential.

$$\boldsymbol{j_2} = \sigma_0 E_0 \begin{pmatrix}\mu^2 B_0^2\cos\theta\\0\\0\end{pmatrix} \quad \text{S 34}$$

To assess the Hall electric field that this creates, we can split $\boldsymbol{j_2}$ up. The first component, $\boldsymbol{j_{2E}}$, is orientated against the original flow of current, reducing the current in the electric field direction. The same thing happens in the disc, and in both cases, this is the first approximation to the magnetoresistance. This component is given by:

$$j_{2E} = \frac{j_2 \cdot \widehat{E}}{|\widehat{E_0}|^2}\widehat{E_0} = \sigma_0 E_0 \begin{pmatrix} \mu^2 B_0^2 \cos^3\theta \\ 0 \\ \mu^2 B_0^2 \sin\theta \cos^2\theta \end{pmatrix} = -\mu^2 B_0^2 \cos^2\theta\, E_0 \qquad \text{S 35}$$

Since it is free to propagate, this component forms the basis of the next iteration. So $j_3$ arises as the deflection of $j_{2E}$, and so on. The other component of $j_2$ we consider is parallel to the surface normal of the cone, $\widehat{u}$. We call this component, $j_{2Norm}$. This is illustrated in Figs S2a-b. Figs S2c-d show the components from the analytic solution for comparison. $\widehat{u}$ is defined as the cross-product of the unit vector in the electric field direction, $\widehat{E}$, and the unit vector in the azimuthal direction, $\widehat{\varphi}$;

$$\widehat{u} = \widehat{E} \times \widehat{\varphi} = \begin{pmatrix} \sin\theta \\ 0 \\ -\cos\theta \end{pmatrix} \qquad \text{S 36}$$

We can then calculate the Hall electric field which is needed to oppose and cancel this current component:

$$E_{2,Hall} = \frac{-j_{2Norm}}{\sigma_0} = -\frac{j_2 \cdot \widehat{u}}{\sigma_0|\widehat{u}|^2}\widehat{u} = \mu^2 B_0^2 E_0 \cos\theta \sin\theta \begin{pmatrix} -\sin\theta \\ 0 \\ \cos\theta \end{pmatrix} \qquad \text{S 37}$$

The next iteration uses only equation S 35:

$$j_3 = \mu(j_{2E} \times B) = \sigma_0 E_0 \begin{pmatrix} 0 \\ \mu^3 B_0^3 \cos^3\theta \\ 0 \end{pmatrix} \qquad \text{S 38}$$

$j_3$ is azimuthal and isn't altered by a Hall potential. Next is $j_4$:

$$j_4 = \mu(j_3 \times B) = \sigma_0 E_0 \begin{pmatrix} -\mu^4 B_0^4 \cos^3\theta \\ 0 \\ 0 \end{pmatrix} \qquad \text{S 39}$$

We see a pattern emerge that every second calculated current component requires projection back onto the cone. The Hall field component of $j_4$ is

$$E_{4,Hall} = \frac{j_2 \cdot \widehat{u}}{\sigma_0|\widehat{u}|^2}\widehat{u} = \sigma_0 E_0 \begin{pmatrix} \mu^4 B_0^4 \cos^3\theta \sin^2\theta \\ 0 \\ -\mu^4 B_0^4 \cos^4\theta \sin\theta \end{pmatrix} = \sigma_0 E_0 \mu^4 B_0^4 \cos^3\theta \sin\theta\, \widehat{u} \qquad \text{S 40}$$

And the electric field component

$$j_{4,E} = \frac{j_4 \cdot \widehat{E_0}}{|\widehat{E_0}|^2}\widehat{E_0} = \sigma_0 E_0 \begin{pmatrix} -\mu^4 B_0^4 \cos^5\theta \\ 0 \\ -\mu^4 B_0^4 \cos^4\theta \sin\theta \end{pmatrix} = \sigma_0 E_0 \mu^4 B_0^4 \cos^4\theta\, E \qquad \text{S 41}$$

Summing the Hall components suggests a full solution for the Hall electric field:

$$E_{Hall} = E_{2,Hall} + E_{4,Hall} \ldots = -\mu^2 B_0^2 E_0 \cos\theta \sin\theta (1 - \mu^2 B_0^2 \cos^2\theta + \mu^4 B_0^4 \cos^4\theta \ldots)\widehat{u} \qquad \text{S 42}$$

Making use again of the series expansion in S 24:

$$\boldsymbol{E}_{Hall} = -\frac{\mu^2 B_0^2 E_0 \cos\theta \sin\theta}{1 + \mu^2 B_0^2 \cos^2\theta} \hat{\boldsymbol{u}} \qquad \text{S 43}$$

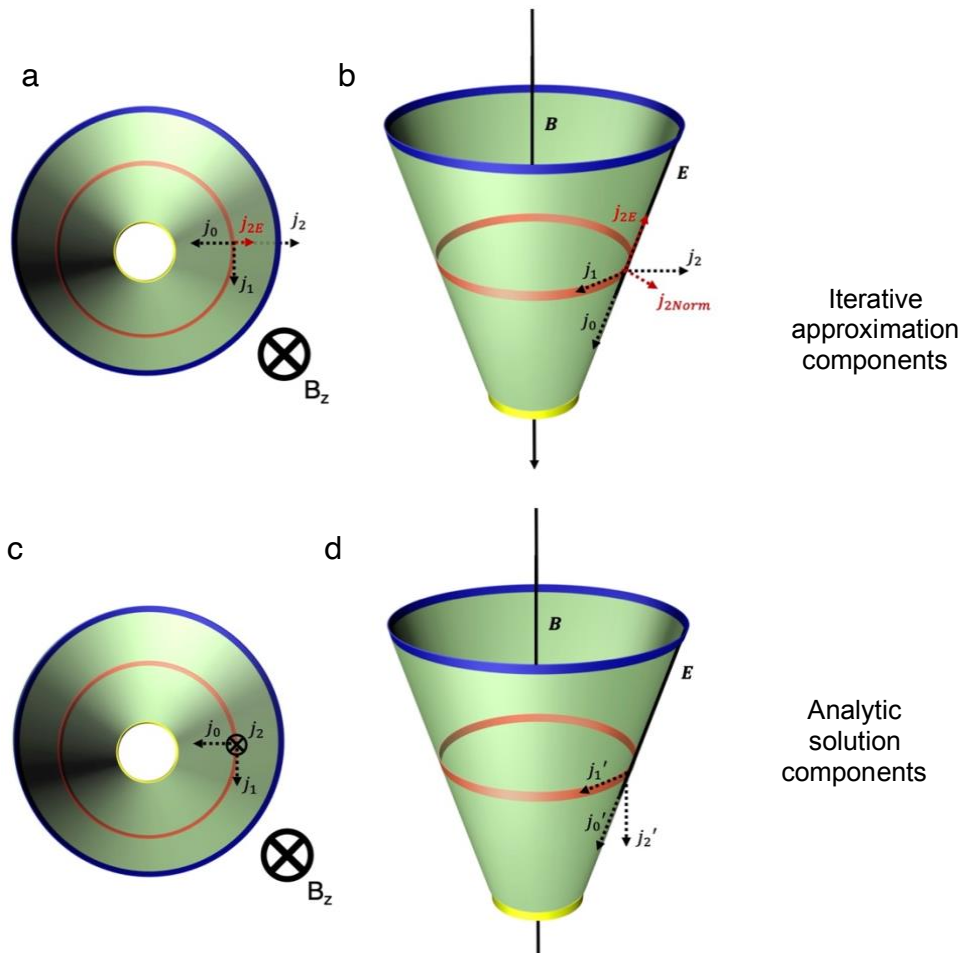

**Figure S2l The Corbino cone.** Plan view (a,c) and oblique view (b,d) of the LNO conducting wall, with the relevant current components as obtained from an iterative procedure (a,b) and analytically (c,d), which converge to the same result.

Key to this argument is the idea that only current components which lie in the cone itself can be allowed to propagate and become deflected. Any other components are eliminated by a Hall field component, as soon as they appear, such that their deflection isn't considered in future iterations. For example, only $\boldsymbol{j}_{2E}$ is used in the calculation of $\boldsymbol{j}_3$, rather than the full component $\boldsymbol{j}_2$.

Finally, we can sum the allowed components to produce the final current density in the cone:

$$\boldsymbol{j}_{total} = \boldsymbol{j}_0 + \boldsymbol{j}_1 + \boldsymbol{j}_{2,E} + \boldsymbol{j}_3 + \boldsymbol{j}_{4,E} \ldots \qquad \text{S 44}$$

$$= \sigma_0 (1 - \mu^2 B_0^2 \cos^2\theta + \mu^4 B_0^4 \cos^4\theta \ldots)[\boldsymbol{E_0} + \begin{pmatrix} 0 \\ \mu B_0 \cos\theta \\ 0 \end{pmatrix}]$$

Which approximates a full solution of:

$$\boldsymbol{j}_{total} = \left[\frac{1}{1 + \mu^2 B_0^2 \cos^2\theta}\right]\sigma_0 \boldsymbol{E_0} + \frac{\sigma_0 E_0}{1 + \mu^2 B_0^2 \cos^2\theta}\begin{pmatrix} 0 \\ \mu B_0 \cos\theta \\ 0 \end{pmatrix} \qquad \text{S 45}$$

Where again the series expansion is used. The magnetoresistance can then be represented as the ratio between the magnitudes of the current components along the external electric field, in the presence and absence of the magnetic field:

$$\frac{j_E(0)}{j_E(B)} = \frac{\sigma_0 E_0}{\frac{1}{1+\mu^2 B_0^2 \cos^2\theta}\sigma_0 E_0} = 1 + \mu^2 B_0^2 \cos^2\theta \qquad \text{S 46}$$

giving a magnetoresistance

$$MR = \frac{j_E(0)}{j_E(B)} - 1 = \mu^2 B_0^2 \cos^2\theta \qquad \text{S 47}$$

as quoted in the main text.

**Section S6: Short test control experiment**

As mentioned in the main text, we completed a control experiment, which tests the external circuit for spurious magnetoresistance. This was done to rule out the possibility that the magnetoresistance we assign to the lithium niobate domain walls is actually an artefact from the external circuit, or indeed arising from the AuCr bottom electrode, which could in principle demonstrate a magnetoresistance. To do this, the MR experiment was repeated with the same sample, save that the LNO film was excluded from the circuit by shorting across to the bottom AuCr electrode. In this way, all other circuit elements can remain the same, with only the LNO film and the connection to the film excluded. We present the results of this short test experiment below in Fig. S3, alongside the equivalent data for the conducting domain walls (Fig. 3c in the main text).

To keep the current consistent between experiments, a series resistor was added in the control experiment, which replicates the resistance of the LNO film in the normal MR experiment. If the LNO is a bystander and the MR signal appears due to some other circuit element in our setup, then we should see an MR signal in this control experiment. Clearly, however, no variation of current with magnetic field exists, and we can hence rule out the notion that the MR is an external artefact.

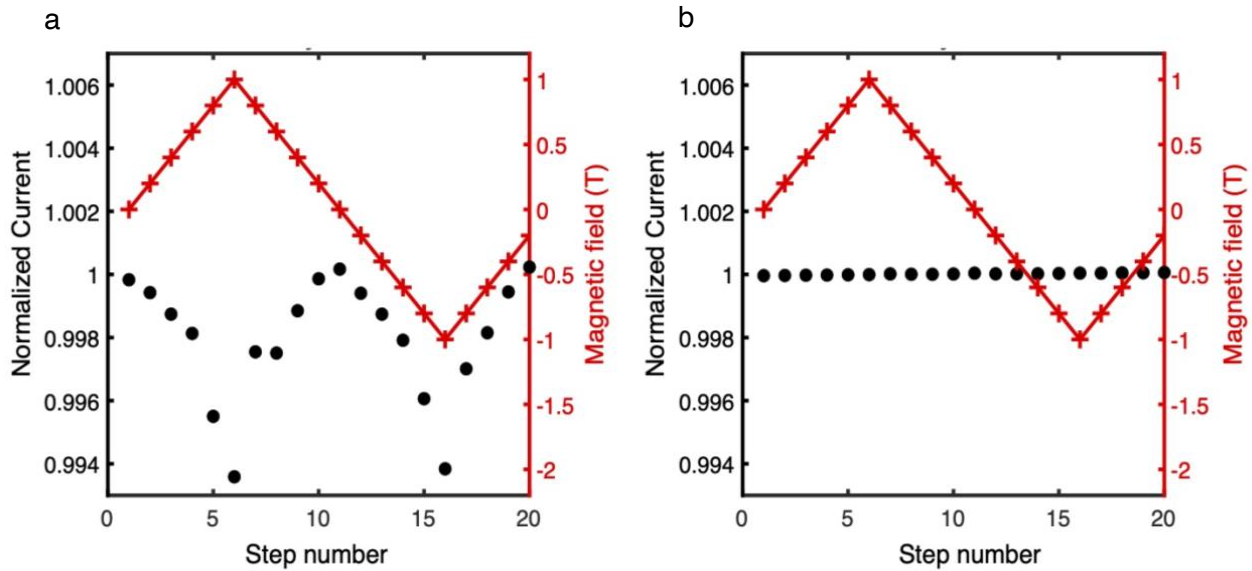

**Figure S3 | Magnetoresistance control experiment. a,** The magnetoresistance response of LNO domain walls, as shown in the main text. The red + show the applied external magnetic field, and the black circles show the normalised current response, averaged over 20 cycles. **b,** The magnetoresistance response during the control experiment. Here, the averaging is done over 13 cycles

**Section S7: Magnetic field influence on domain structure**

To ensure the magnetic field has no effect on the domain structure, we performed PFM mapping in a magnetic field. First, conducting domain walls were injected into the as-received LNO, using an AFM tip as the top electrode. PFM domain mapping was carried out before and after magnetic field cycling, which was done using the same magnet as in the MR measurements (Fig. S4). No changes in domain structure were seen.

Then, PFM was performed while a magnetic field was applied to the sample in situ (Fig. S5). The AFM magnetic field unit is designed to apply an in-plane magnetic field to the system (Fig. S5a). No changes to the domain structure were seen with the field in this in-plane geometry (Fig. S5b-d). We accessed the out-of-plane geometry by moving the sample to a spot on the magnetic platen where an out-of-plane component of the field is expected, as per the AFM manual (illustrated in Fig S5e, taken from Asylum Research MFD infinity User manual). Note that this magnitude of the out-of-plane component will likely not match that measured by the Hall probe in the system (quoted in the figure titles in Fig. S5f-h). Again, no changes to the domain structure were seen in the presence of any magnetic field.

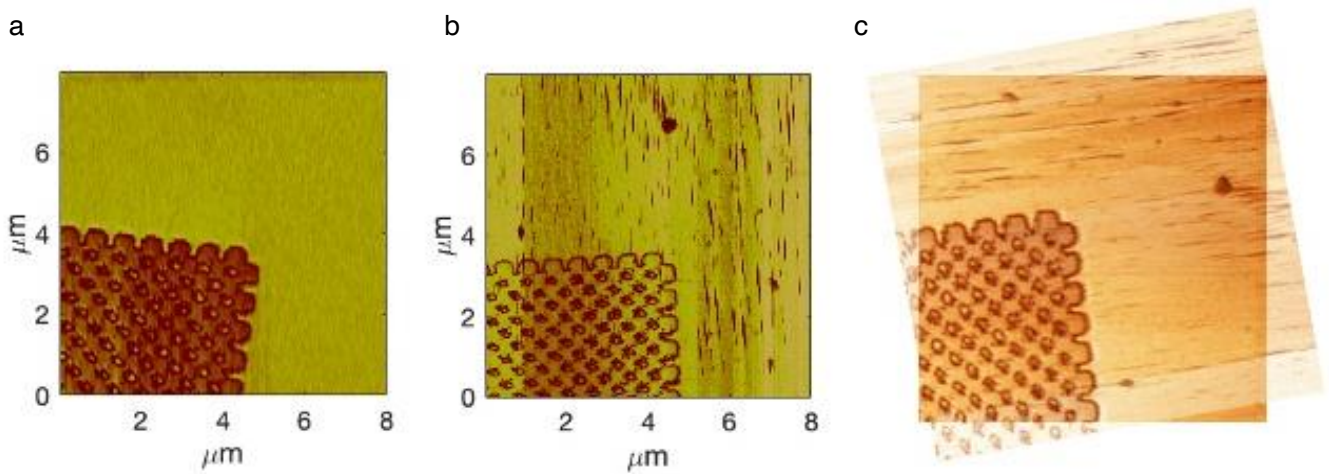

**Figure S4 | Magnetic field influence on domain structure. a,** PFM amplitude showing the domain structure in LNO before application of a magnetic field cycle similar to that applied in the MR experiment. **b,** PFM amplitude after magnetic field cycle. **c,** An image showing the overlap of PFM pre and post magnetic field cycle. The colour scheme has been changed for clarity.

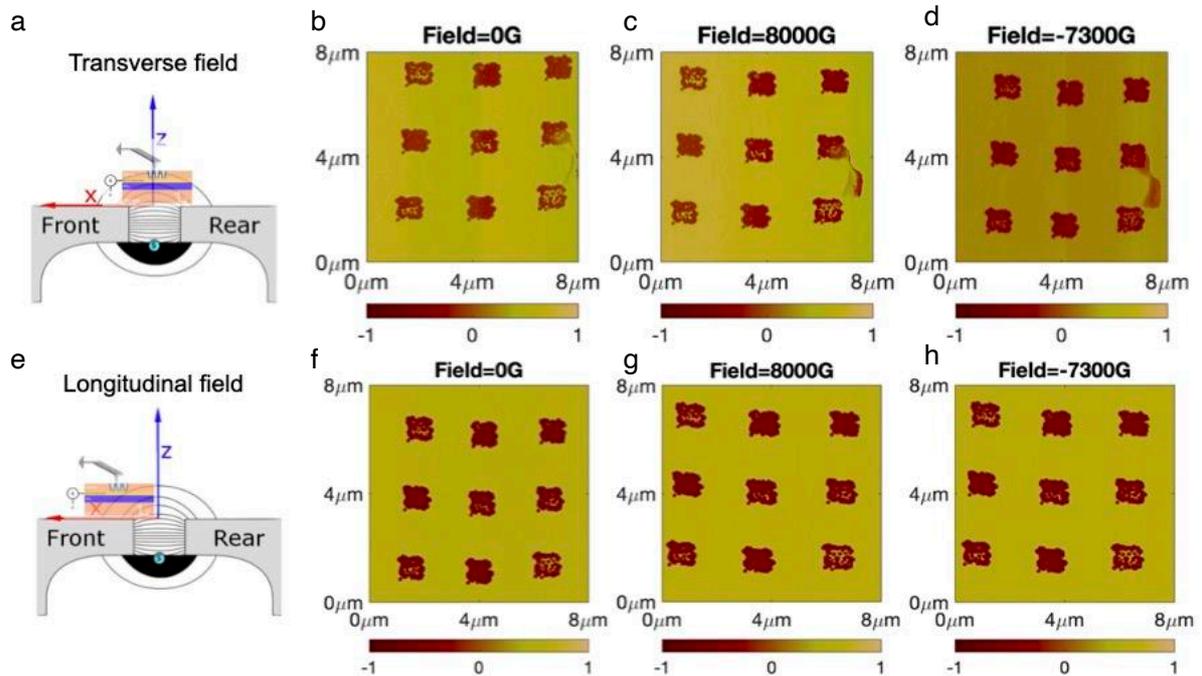

**Figure S5 | Magnetic field influence on domain structure in situ. a,** Illustration of the field profile from an AFM manual (MFD infinity User manual), illustrating the magnetic field profile for the in-situ AFM magnet. The sample experiences a transverse field. **b-d,** PFM of conducting domain walls at 0G, +8000G and -7300G respectively, in a transverse geometry. **e,** an illustration of the LiNbO$_3$ sample moved slightly, so as to experience a slight longitudinal component of magnetic field. **f-h,** PFM of conducting domain walls at 0G, +8000G and -7300G respectively, in a the longitudinal geometry. Note that here the quoted field (in the subplot titles) is likely higher than the longitudinal field experienced by the sample.

**Section S8: IV characteristics of the LNO samples and the possibility of metallicity**

We show the current-voltage (IV) characteristics of the LNO samples in Fig. S6. Fig. S6a illustrates the "switching" pulse applied to the fresh LNO surface to induce the domain walls. A 100kOhm resistor was set in series during switching to limit the current through the system. IV performed before and after application of the switching pulse (Fig. S6b) shows clearly a marked and persistent increase in current through the device spanning 4-5 orders of magnitude, confirming the existence of conducting domain walls between the top and bottom electrode. Fig. S6c shows the temperature dependence of IV for a sample with domain walls injected. IV curves were performed at room temperature and then again at various temperatures upon ramping up to a maximum temperature of 80°C (solid lines). IV data was also gathered at the same temperatures on the ramp back down to room temperature (dotted lines). Several things are visible in Fig. S6c:

- Firstly, at a given temperature, the current at a set voltage increases slightly upon application of successive IV loops. This property is well known – the LNO domain wall conductivity state can be changed by application of a sub-coercive bias[2]. This behaviour has been investigated for potential memristive device applications[3].
- Secondly, current clearly increases with increasing temperature. This suggests a semiconducting response; however, we note that these domain wall conductance measurements (along with most others reported in literature) are 2-probe, meaning the response of the electrode contact is convoluted with any real domain wall response.
- Thirdly, there appears to be some hysteresis in the temperature cycle; the current appears to be diminished in the IV curves performed on the ramp "back down" from high temperature. A decay of domain wall conductivity with time is well documented in LNO domain walls and elevated temperature has been seen to accelerate the decay process. Several mechanisms have been proposed to explain the decay: cross-sectional TEM images have shown the domain wall tilt angle relaxing after some time[2]. This process could, in principle, happen quicker at higher temperatures. It has also been suggested that ion migration occurs, screening the domain wall conductivity. This is supported by the observed fact that conductivity decay appears to be "activated"[4]. We note here again that all conclusions on this matter are also subject to parasitic contact effects, which highlights the need for genuine 4-probe transport measurements on these systems to reveal reliable transport phenomena.

While our IV measurements demonstrate a semiconducting temperature dependence, we reiterate that these measurements, like most performed on domain walls, are 2-probe. The underlying behaviour of the domain wall system could be very different from that suggested in a 2-probe measurement, with the electrode-sample contact playing a dominant role, obscuring the inherent domain wall response. Given the huge change in conductivity demonstrated in these systems, it is interesting to consider, at least theoretically, the possibility of metallic conduction along the domain walls. After all, hints of metallicity have been seen in other domain wall systems (such as barium[5] and lead[6] titanate). Low carrier densities in the LNO domain walls might initially suggest that the Mott criterion for an insulating-to-metal transition should be unlikely, but further consideration shows this is a viable possibility: the conditions needed for a Mott transition are encapsulated in a simple criterion

$$n_c^{1/3} a_h^* \leq 0.25 \qquad \text{S 48}$$

where $n_c$ is the critical carrier density and $a_h^*$ the effective Bohr radius of the electron-centre system. Semiconductors with low effective mass have higher Bohr radii and require a low density of carriers to achieve this transition. If the high mobility that we measure is indicative of a low effective mass, the critical density for metallicity could be low; perhaps low enough to be surpassed by the local change in carrier density at the wall which occurs to screen the bound polarisation charge. Using a rough estimate for the Bohr radius in LNO

$$a_h^* = \frac{K_{st} \hbar^2}{m^* e^2} = 7 \times 10^{-9} m \qquad \text{S 49}$$

where, $K_{st}$ is the low frequency static dielectric constant ($K_{st} \sim 84\varepsilon_0$)[7], and $m^*$ is the effective mass of the electron. This is not measured, but an estimate is suggested as $m^* \sim 0.05 m_e$[8]. The critical density for an insulator-metal transition is then:

$$n_c \approx 4 \times 10^{16} cm^{-3} \qquad \text{S 50}$$

It is interesting to note that the domain wall width required to surpass this critical density for metallicity in LNO walls is 0.25nm (using eqn 10 in the main text). While domain wall widths are usually assumed to be in the nanometer range, our estimate is not far off this value. Combined with the rough, order of magnitude nature of the preceding discussion, this analysis suggests metallic transport is a viable possibility for LNO walls. For reference, a wall width of 0.25nm, along with the estimated active carrier density, results in a conductivity of the walls of:

$$\sigma_{DW} \approx 2400 \ (\Omega m)^{-1} \qquad \text{S 51}$$

This is higher than previously reported estimates[4]. However, the domain wall is assumed to be thinner here, and the domain wall inclination angle with respect to the polar axis, which is known to affect wall conduction[9], is also higher in our samples.

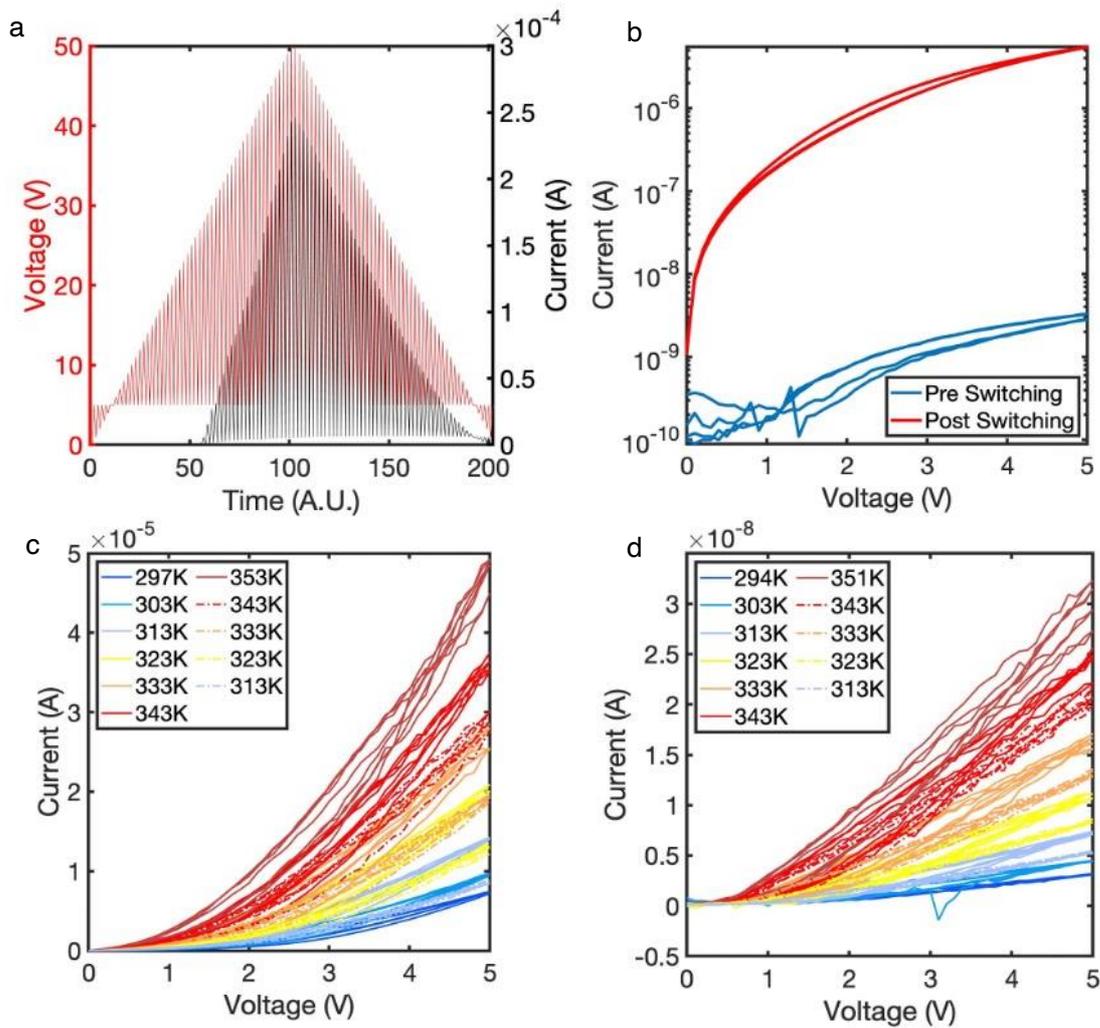

**Figure S6 | IV response and temeperature dependent IV response of MR LiNbO$_3$ samples. a,** The switching pulse, and measured current, applied to the LNO sample to induce domain walls. **b,** IV before and after switching, showing a clear increase in sample conductance **c,** Temperature variation of the IV response of the LNO sample with domain walls induced. Full lines indicate the IV was taken on the ramp up in temperature, dotted lines on the ramp down. **d,** temperature dependent IV of the sample before inducing domain walls, showing the response of the fresh LNO film.

**Section S9: MR as a function of field angle**

As reported in the main text, we have repeated the MR experiment for a series of magnetic field angles and found no clear variation of the MR (main text Fig. 4b). If anything, the MR increases as the field orientation changes from being parallel to being perpendicular to the conical axes of the Corbino cones. Here we lay out the quantitative derivation for geometric MR with the field applied in the XY plane (hereafter XY-MR). It shows that an XY-MR of similar magnitude to the z-oriented field case is expected (hence the apparent isotropy in the MR response that we measure). The reason MR exists in this field orientation is because the charged "boundaries" that develop do not sustain the full Hall field required to fully cancel Lorentz deflection of carriers. This is a well-known result of current deflection in samples with "intermediate" length to width ratios[10] (Fig. S7c), which are neither long Hall bars ($l/w > 5$, Fig S7a) nor a sufficiently short Hall bar ($l/w \ll 1$, shown in Fig. S7b). Note the limiting case $l/w = 0$ is the Corbino disc.).

First, we discuss the domain wall geometry in detail, estimating the domain wall dimensions, based on microstructural measurements from the main text. This allows us to approximate the effective aspect ratio in our samples. Then, we solve the current density equations for a magnetic field in the x-y plane by Popovic's iterative method, taking care to note than some current components will be partially reduced by a Hall field, with the degree of cancellation governed by a geometry factor. Finally, we derive the form of the XY-MR in terms of this geometry factor. Using estimates for it from literature, we find the geometric XY-MR to be of a similar magnitude to that expected when the magnetic field is parallel to the z-axis, and therefore similar in magnitude to that experimentally measured. This demonstrates that the interpretation of MR as having a geometric origin is fully consistent with all of our experimental observations and that hence the mobility values inferred are likely to be robust. Certainly, there is no need to evoke any mechanism beyond geometric MR to fully explain our data.

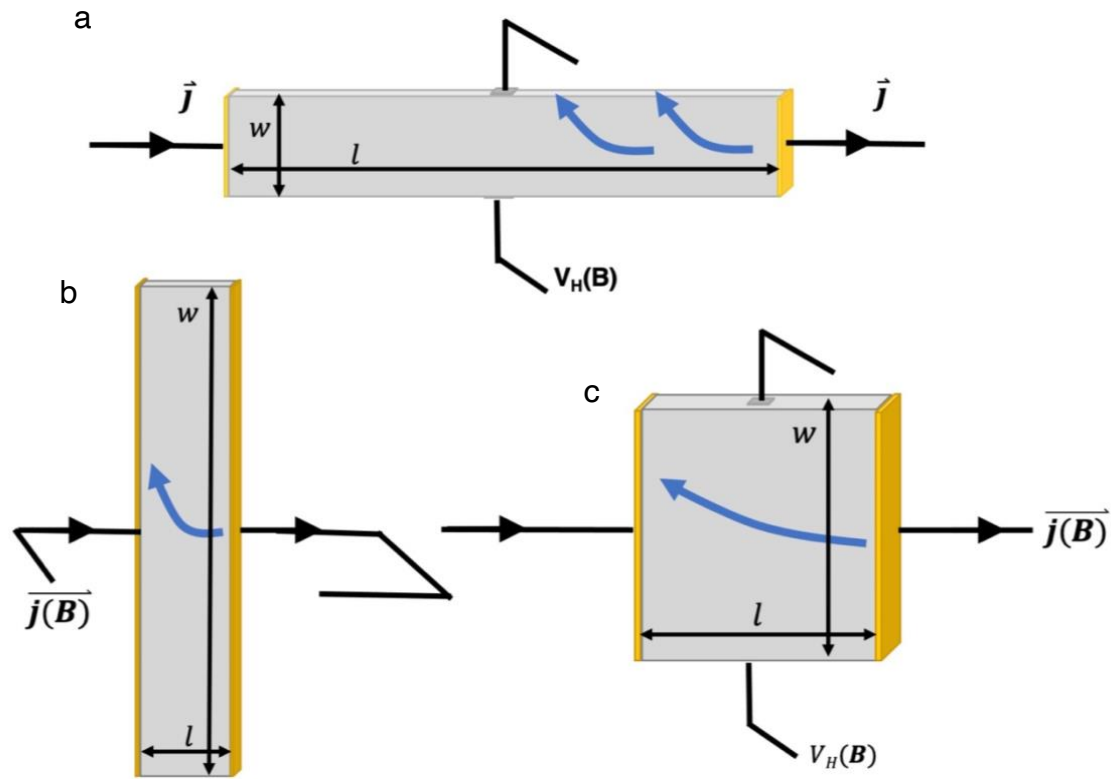

**Fig S7| Sample Geometries. a,** A standard long Hall bar, with $l/w > 5$. **b,** A short hall bar, which seeks to maximise the geometric magnetoresistance effect: $l/w \ll 1$. **c,** An intermediate sample, with $1 < l/w < 5$. Here, both the Hall effect and geometric MR are present.

**Domain wall geometry**
Taking the radius of the conical walls at the top electrode as $r_1 = 150$nm, and the wall angle away from the thin film normal as 90-$\theta$ = 12° (where $\theta$ is the inclination angle of the walls as defined in the main text) over a film of thickness $t = 500$nm, we find a bottom radius of

$$r_2 = r_1 - t\tan(90 - \theta) = 44nm \qquad \text{S52}$$

as illustrated in Fig. S8a. We can "unwrap" the conducting conical surface to consider it as a 2D truncated sector of a circle, with upper and lower half arc-widths of $\pi r_1$ and $\pi r_2$, separated by a fixed length ($l$):

$$l = \frac{t}{\cos(90 - \theta)} = 511 nm \qquad \text{S53}$$

as seen in Fig. S8b. Given the shape of the unwrapped cone, the half arc-width of the conductor changes as we traverse from the top to the bottom electrode contacts. Therefore, we define a variable coordinate $\rho$, which maps the perpendicular distance of any point on the domain wall surface from the top electrode. The half arc-width of the conducting cone at any point $\rho$ is simply half of the circumference of the circular section of the cone at that point. With the help of figure S8a, we can see that the radius of this circular section, as a function of $\rho$ becomes:

$$r(\rho) = r_1 - x'(\rho) = r_1 - \rho \sin(90 - \theta) \qquad \text{S54}$$

The half arc-width ($w(\rho)$) is therefore:

$$w(\rho) = \pi r(\rho) = \pi(r_1 - \rho \sin(90 - \theta)) \qquad \text{S55}$$

Figure S8c shows a plot of $l/w(\rho)$ as a function of the normalised coordinate $\rho/l$, illustrating the aspect ratios involved in our system and how they change along the current path. Half arc-widths have been defined because the Lorentz force associated with XY-MR geometry creates diametrically opposite strips of increased and decreased carrier density (illustrated in figure S9b and S9c). As a result, two symmetrically related regions form which act as short Hall bars, electrically connected in parallel. Ideally, the Hall electric field is azimuthal and so flux lines follow arcs in the 2D representation of the cone, always being perpendicular to the undeflected current direction. The centres of charge density are separated by the width defined in equation S55, such that this half arc-width is the effective width of the Hall bar. We note that the width of the Hall bar varies along the length of the current path.

After ref [11,12], we can also define the "geometrical factor of magnetoresistance", which is the ratio between the measured geometric MR in a sample of intermediate $l/w$ and the geometric magnetoresistance that would be measured in a perfect Corbino disc ($l/w = 0$)

$$g_{MR}(l/w) = \frac{MR(l/w)}{MR(0)} \qquad \text{S52}$$

$g_{MR}$ is 1 for a Corbino disc ($l/w = 0$), and 0 for an infinitely long Hall bar. Physically, $g_{MR}$ tracks how much the shorting of the Hall potential, in samples with intermediate aspect ratios, leads to carrier path bending, and therefore geometric MR. The relative change in transverse resistance can be written as

$$r(B) = \frac{R(B)}{R(0)} = \frac{\sigma(0)}{\sigma(B)}(1 + \mu^2 B^2 g(l/w)) \qquad \text{S53}$$

Where the first term accounts for physical MR and the second term the geometric MR. For the reasons given in the main text, we assume that the physical MR is insignificant in comparison to the geometric MR term (the usual case when the carrier mobility is reasonably high in comparison to mobility spread). Equations S56 and S57 are valid for $\mu B < 1$. Considering the geometrical MR contribution only, we then find a general magnetoresistance of

$$MR = \frac{R(B)}{R(0)} - 1 = \mu^2 B^2 g(l/w) \qquad \text{S54}$$

in intermediate "short-Hall-bar" geometries.

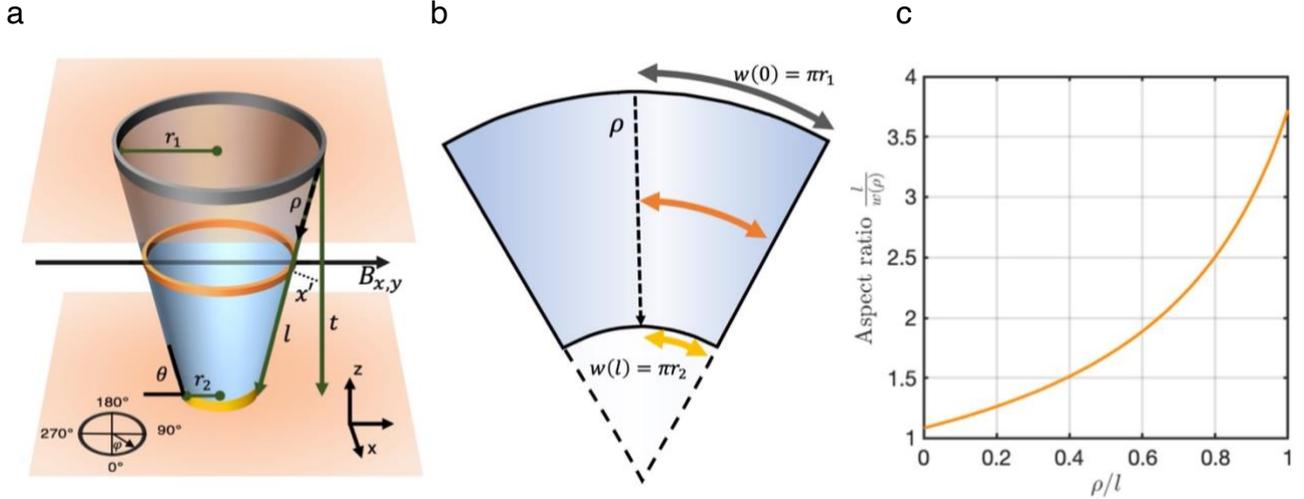

**Fig. S8: Domain wall dimensions. a**, Schematic of the domain wall cone in the XY-MR configuration with some dimensions labelled. The coloured rings are circular sections of the cone at different depths into the LNO film. **b,** schematic of the cone surface "unwrapped", illustrating the truncated circle sector geometry. Dotted lines represent variables defined in the text above and the solid, double-ended, arrows represent the half arc-widths corresponding to half the circumference of the rings in **a**. **c,** the aspect ratio of the conductor shown in **b**, as a function of normalised length along the current path ($\rho/l$). The range of aspect ratios are consistent with it being considered as a short Hall bar at all points.

**Solution to the current density equations:** Here we solve the current density equations for current in a conical geometry when exposed to specific electric and magnetic fields $B$ and $E$. As always, the high symmetry axis of the cone is taken to be aligned along the z-axis, but this time the magnetic field is applied in the x-y plane. We choose the y axis for convenience, but the behaviour is the same as for any other in-plane direction by symmetry. We begin by setting out the field profiles for the solution. In cylindrical base coordinates, the electric and magnetic fields are:

$$\boldsymbol{E} = \begin{pmatrix} E_r \\ E_\varphi \\ E_z \end{pmatrix} = E_0 \begin{pmatrix} -\cos\theta \\ 0 \\ -\sin\theta \end{pmatrix}; \quad \text{S55}$$

$$\boldsymbol{B} = \begin{pmatrix} B_r \\ B_\varphi \\ B_z \end{pmatrix} = T_{ij} \begin{pmatrix} B_x \\ B_y \\ B_z \end{pmatrix} = \begin{pmatrix} \cos\varphi & \sin\varphi & 0 \\ -\sin\varphi & \cos\varphi & 0 \\ 0 & 0 & 1 \end{pmatrix} \begin{pmatrix} 0 \\ B_0 \\ 0 \end{pmatrix} = \begin{pmatrix} B_0 \sin\varphi \\ B_0 \cos\varphi \\ 0 \end{pmatrix} \quad \text{S56}$$

Here, $\boldsymbol{E}$ and $\boldsymbol{B}$ are electric and magnetic field vectors, $r, \varphi$ and $z$ subscripts represent radial, azimuthal and z components of the vectors, and $\theta$ is the inclination angle of the cone away from the horizontal (approx. 78°). $T_{ij}$ is the matrix used to transform vectors from Cartesian space into cylindrical space. $\varphi$, when not used in a subscript, represents the azimuthal coordinate.

The calculation of the current density proceeds similarly to that in the Z-MR treatment; the effect of the magnetic field is found by iteratively summing smaller and smaller corrections to the drift current, to successively higher orders of $\mu B$. We split each appearing current component into sub-components, which are either embedded in the conducting cone (and allowed to propagate) or directed off the conical surface (to then be cancelled by a Hall potential developed across the width of the domain wall). To do this, we need the unit vector in the direction of the cone surface normal. The electric field, $\widehat{E}$ lies within the cone. Furthermore, since a cone has rotational symmetry, a translation of $\widehat{\varphi}$ also lies within the cone, such that the vector

$$\hat{\boldsymbol{u}} = \hat{\boldsymbol{E}} \times \hat{\boldsymbol{\varphi}} = \begin{pmatrix} sin\theta \\ 0 \\ -cos\theta \end{pmatrix} \quad \text{S61}$$

will always be perpendicular to the tangent plane on the cone surface.

We first calculate the drift due to the electric field only:

$$\boldsymbol{j}_0 = \sigma \boldsymbol{E} = \sigma E_0 \begin{pmatrix} -cos\theta \\ 0 \\ -sin\theta \end{pmatrix} \quad \text{S62}$$

Then, using the same process as before (equation S19, as used in the Z-MR treatment), we calculate the "first" deflected current term, $\boldsymbol{j}_1$,

$$\boldsymbol{j}_1 = \mu(\boldsymbol{j}_0 \times \boldsymbol{B}) = \sigma \mu E_0 \begin{pmatrix} B_0.sin\theta.cos\varphi \\ -B_0.sin\theta.sin\varphi \\ -B_0.cos\theta.cos\varphi \end{pmatrix} \quad \text{S63}$$

In contrast to the Z-MR case, where the first deflected current term is azimuthal (and therefore entirely within the conical surface), this term needs to be split into subcomponents, as the vector does not lie in the conducting cone surface everywhere. First, the "Hall" subcomponent is found by taking the projection of the current along the surface normal ($\hat{\boldsymbol{u}}$), and multiplying by -1 (because a Hall component will act to *oppose* this surface normal current)

$$\boldsymbol{j}_{1,H} = -\frac{\boldsymbol{j}_1 \cdot \hat{\boldsymbol{u}}}{|\hat{\boldsymbol{u}}|^2}\hat{\boldsymbol{u}} = -\sigma \mu E_0 B_0 cos\varphi\, \hat{\boldsymbol{u}} \quad \text{S64}$$

Then, the remaining "allowed" current subcomponent is found by adding the total current component and the Hall subcomponent corrections:

$$\boldsymbol{j}_{1,All} = \boldsymbol{j}_1 + \boldsymbol{j}_{1,H} = \sigma \mu E_0 \begin{pmatrix} 0 \\ -B_0.sin\theta.sin\varphi \\ 0 \end{pmatrix} \quad \text{S65}$$

We find that only an azimuthal term now survives. Inspecting equation S65, we see that the deflection of carriers now depends on the azimuthal coordinate, $\varphi$. In other words, the magnitude and sense of the deflection depends on the angular location on the conical surface. We can rationalize this with the help of figure S9a. At the points in the cone corresponding to $\varphi = 0°$ and $180°$, the deflection takes us directly off the cone, and so should be fully compensated by a Hall field across the domain wall width. Conversely, this deflection remains fully within the cone at points $\varphi = 90°, 270°$, leading to maximal allowed azimuthal current components at these points. $\boldsymbol{j}_{1,All}$ is illustrated by red arrows in fig S9a. This explains the $sin\varphi$ dependence in the first current component.

In contrast to the Z-MR treatment (where both the original system and the perturbing magnetic field had rotational symmetry about the z axis), we now have a situation where the rotational symmetry is broken by the magnetic field. Therefore, we might expect that properties such as carrier density, current density etc will vary with azimuthal angle $\varphi$. If we look at the deflection implied by equation S65, we find that there is a preference for carriers to deflect towards one side of the cone (Fig. S9a), which should lead to the formation of a new Hall potential (Fig. S9b) distinct from that present across the width of the domain wall itself.

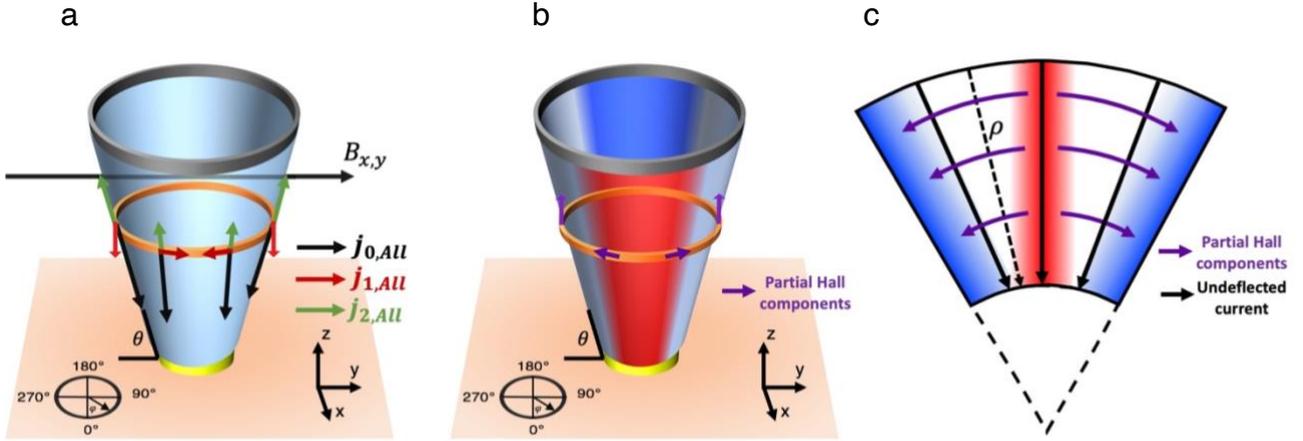

**Fig S9: Current components with magnetic field in the x-y plane a,** Current components obtained from the iterative solution. **b,** The building of a Hall potential across the conducting cone itself. Red and blue shaded areas indicate the accumulation/depletion of carriers, and purple arrows represent the effective current density which would result from the associated partial Hall potential difference developed across this effective pair of short Hall bars. **c,** the unwrapped cone, with red and blue regions highlighting the building of the Hall potential across the half arc widths. Purple arrows represent the azimuthal Hall flux lines, and the black solid arrows represent the undeflected current. The dotted black line represents the variable distance $\rho$.

We need to assess the degree to which this new Hall potential will cancel the Lorentz deflection of carriers for this component. As mentioned above, this comes down to the geometry of the situation; specifically, the length-to-width ratio of the current channel, and the term $g_{MR}(l/w)$ in equation S56. Here, length is the distance between current carrying electrodes, and width is the arc length distance between the strips where charge accumulation / denudation occurs (red and blue strips in figure S9b). Figure S9c shows the cone unwrapped into a truncated circular sector, illustrating the Hall potential more clearly. As we can see, the red and blue strips are separated by half the circumference of the circular section of the cone at all points $\rho$. The Hall electric field will act azimuthally, along the circular arcs of the conducting cone (purple arrows), to counteract the deflection of current due to the magnetic part of the Lorentz force. This cancellation is not total, however, because the geometry reflects that of a short Hall bar. We therefore assume that the new fractional current, allowed in the $\widehat{\varphi}$ direction, is simply the full allowed current density, multiplied by the geometry factor $g_{MR}$. Essentially, Hall fields now act to cancel all of the $\widehat{u}$ oriented current, and some fraction of the $\widehat{\varphi}$ oriented current, where the fraction is determined by $g_{MR}$:

$$\boldsymbol{j}_{1,All} = \boldsymbol{j}_1 + \boldsymbol{j}_{1,H} = \boldsymbol{j}_1 - \frac{\boldsymbol{j}_1 \cdot \widehat{\boldsymbol{u}}}{|\widehat{\boldsymbol{u}}|^2}\widehat{\boldsymbol{u}} - G_H \frac{\boldsymbol{j}_1 \cdot \widehat{\boldsymbol{\varphi}}}{|\widehat{\boldsymbol{\varphi}}|^2}\widehat{\boldsymbol{\varphi}} \qquad \text{S66}$$

Here, $G_H$ determines the amount of current along the $\widehat{\theta}$ direction which is cancelled by the partial Hall potential, such that

$$1 - G_H = g_{MR} \qquad \text{S67}$$

Evaluating this, we find a current density

$$\boldsymbol{j}_{1,All} = \sigma\mu E_0 B_0 \sin\theta \sin\varphi (1 - G_H)\begin{pmatrix}0\\-1\\0\end{pmatrix} \qquad \text{S68}$$

Continuing the iterative approach, the next current component is:

$$\boldsymbol{j}_2 = \mu(\boldsymbol{j}_{1,All} \times \boldsymbol{B}) = \sigma E_0 \mu^2 B_0^2 (1 - G_H) \begin{pmatrix} 0 \\ 0 \\ \sin\varphi^2 \sin\theta \end{pmatrix} \quad \text{S69}$$

Again, we project this component along $\hat{\boldsymbol{u}}$ and $\hat{\boldsymbol{\varphi}}$ to find the appropriate Hall fields and resulting allowed current:

$$\boldsymbol{j}_{2,All} = \boldsymbol{j}_2 - \frac{\boldsymbol{j}_2 \cdot \hat{\boldsymbol{u}}}{|\hat{\boldsymbol{u}}|^2} \hat{\boldsymbol{u}} - G_H \frac{\boldsymbol{j}_2 \cdot \hat{\boldsymbol{\varphi}}}{|\hat{\boldsymbol{\varphi}}|^2} \hat{\boldsymbol{\varphi}} \quad \text{S70}$$

Explicitly:

$$\boldsymbol{j}_{2,All} = \sigma E_0 \mu^2 B_0^2 (1 - G_H) \begin{pmatrix} 0 \\ 0 \\ \sin\varphi^2 \sin\theta \end{pmatrix} - (-\mu^2 B_0^2 \sigma E_0 \sin^2\varphi \sin\theta \cos\theta)(1 - G_H) \begin{pmatrix} \sin\theta \\ 0 \\ -\cos\theta \end{pmatrix} - G_H \begin{pmatrix} 0 \\ 0 \\ 0 \end{pmatrix} \quad \text{S71}$$

Which we can write in terms of the external electric field

$$\boldsymbol{j}_{2,All} = -\mu^2 B_0^2 \sigma E_0 \sin^2\varphi \sin^2\theta\ (1 - G_H) \begin{pmatrix} -\cos\theta \\ 0 \\ -\sin\theta \end{pmatrix} = -\alpha \sigma \boldsymbol{E} \quad \text{S72}$$

where

$$\alpha = \mu^2 B_0^2 \sin^2\theta \cdot \sin^2\varphi (1 - G_H) \quad \text{S73}$$

Next:

$$\boldsymbol{j}_3 = \mu(\boldsymbol{j}_{2,All} \times \boldsymbol{B}) = -\mu^3 B_0^3 \sigma E_0 \sin^2\varphi \sin^2\theta\ (1 - G_H) \begin{pmatrix} \sin\theta \cdot \cos\varphi \\ -\sin\theta \cdot \sin\varphi \\ -\cos\theta \cdot \cos\varphi \end{pmatrix} \quad \text{S74}$$

and again, we find the allowed component:

$$\boldsymbol{j}_{3,All} = \boldsymbol{j}_3 - \frac{\boldsymbol{j}_3 \cdot \hat{\boldsymbol{u}}}{|\hat{\boldsymbol{u}}|^2} \hat{\boldsymbol{u}} - G_H \frac{\boldsymbol{j}_3 \cdot \hat{\boldsymbol{\varphi}}}{|\hat{\boldsymbol{\varphi}}|^2} \hat{\boldsymbol{\varphi}} = \mu^3 B_0^3 \sigma E_0 \sin^3\varphi \sin^3\theta\ (1 - G_H)^2 \begin{pmatrix} 0 \\ 1 \\ 0 \end{pmatrix}$$
$$= \alpha \sigma \mu E_0 B_0 \sin\theta \sin\varphi (1 - G_H) \begin{pmatrix} 0 \\ 1 \\ 0 \end{pmatrix} \quad \text{S75}$$

The next component is

$$\boldsymbol{j}_4 = \mu^4 B_0^4 \sigma E_0 \sin^3\varphi \sin^3\theta\ (1 - G_H)^2 \begin{pmatrix} 0 \\ 0 \\ -\sin\varphi \end{pmatrix} \quad \text{S76}$$

giving the allowed components:

$$\boldsymbol{j}_{4,All} = \mu^4 B_0^4 \sigma E_0 \sin^4\varphi \sin^4\theta\ (1 - G_H)^2 \begin{pmatrix} -\cos\theta \\ 0 \\ -\sin\theta \end{pmatrix} = \alpha^2 \sigma \boldsymbol{E} \quad \text{S77}$$

One more current iteration allows us to see the pattern clearly:

$$\boldsymbol{j}_5 = \mu(\boldsymbol{j}_{4,All} \times \boldsymbol{B}) = \mu^5 B_0^5 \sigma E_0 \sin^4\varphi \sin^4\theta\ (1 - G_H)^2 \begin{pmatrix} \sin\theta \cdot \cos\varphi \\ -\sin\theta \cdot \sin\varphi \\ -\cos\theta \cdot \cos\varphi \end{pmatrix} \quad \text{S78}$$

and

$$\boldsymbol{j}_{5,All} = -\mu^5 B_0^5 \sigma E_0 \sin^5\varphi \sin^5\theta\, (1-G_H)^3 \begin{pmatrix} 0 \\ 1 \\ 0 \end{pmatrix} = -\alpha^2 \sigma\mu E_0 B_0 \sin\theta\, \sin\varphi(1-G_H) \begin{pmatrix} 0 \\ 1 \\ 0 \end{pmatrix} \quad \text{S79}$$

Summing the allowed components, we find a current density of

$$\boldsymbol{j}_{total} = \sum_{n=0}^{\infty} \boldsymbol{j}_{n,All} = (1 - \alpha + \alpha^2 - \alpha^3 \dots)\sigma\left(\boldsymbol{E} + E_0\begin{pmatrix} 0 \\ -\mu B_0 \sin\theta\, \sin\varphi(1-G_H) \\ 0 \end{pmatrix}\right) \quad \text{S80}$$

Which generates a full solution:

$$\boldsymbol{j}_{total} = \frac{1}{1+\alpha}\left(\sigma\boldsymbol{E} + \sigma E_0\begin{pmatrix} 0 \\ -\mu B_0 \sin\theta\, \sin\varphi(1-G_H) \\ 0 \end{pmatrix}\right) \quad \text{S81}$$

In a similar way to the z-directed magnetic field case, our magnetoresistance is given by

$$MR(\varphi) = \alpha = \mu^2 B_0^2 \sin^2\theta \cdot \sin^2\varphi (1-G_H) \quad \text{S82}$$

Integrating over the entire cone, to remove the $\varphi$ dependence and obtain MR, in terms of current (rather than current density)

$$MR = \frac{1}{2\pi}\int_0^{2\pi} \mu^2 B_0^2 \sin^2\theta \cdot \sin^2\varphi(1-G_H)d\varphi = \frac{\mu^2 B_0^2 \sin^2\theta\, g_{MR}}{2} \quad \text{S83}$$

where the substitution in S67 has been made.

We can turn to literature, in order to estimate $g_{MR}$. In rectangular plates, Lippmann and Kuhrt[11] first derived $g_{MR}$ for the full aspect ratio-Hall angle parameter space. In regions where both the Hall angle is small ($\tan(\Theta) = \mu B \approx \Theta \ll 1$) and where $\frac{l}{w} > 1$, equations S56-S58 are valid, with the geometric factor of magnetoresistance approximated by

$$g(l/w) = \frac{14}{\pi^3} S_3 \frac{w}{l} \quad \text{S84}$$

where $S_3$ is a constant approximately equal to 1.2. Using this form of the geometric factor, the XY-MR can be calculated across the entire range of aspect ratios that our domain walls exhibit and the average geometric factor used as a measure of how our system will behave. This is evaluated by integrating the function in equation S84 with respect to $\gamma = l/w$ between the limiting aspect ratios demonstrated in figure S8c ($\gamma_{min} = 1.09, \gamma_{max} = 3.72$). We find the average geometric factor for our system is:

$$g(l/w)_{ave} = \frac{1}{\gamma_{max} - \gamma_{min}}\int_{\gamma_{min}}^{\gamma_{max}} \frac{14}{\pi^3} S_3 \frac{1}{\gamma}\, d(\gamma) = \frac{1}{2.63}\frac{14}{\pi^3}S_3 \ln\left(\frac{3.72}{1.09}\right) = 0.25 \quad \text{S85}$$

We therefore find a magnetoresistance of:

$$MR_{xy} = \frac{0.25\, \mu^2 B_0^2 \sin^2\theta}{2} \quad \text{S86}$$

From our measurements presented in the main text, $MR_{xy} \sim 0.4 \times 10^{-2}$ at 1T, but we think this is slightly suppressed by series contact resistance ($MR_{xy} \sim 0.6 \times 10^{-2}$ at 1T might be closer to the true value, as this was the Z-MR obtained with lower resistance contacts). Using equation S86, to estimate the carrier mobility implied, gives (to 1 significant figure) a value of 2,000cm$^2$V$^{-1}$s$^{-1}$. Another way of probing the physics implied is to use the mobility value estimated from our Z-MR measurements (3,700cm$^2$V$^{-1}$s$^{-1}$) to see how large the $MR_{xy}$ could be. Doing this, we find values in the order of 1% at 1T - more than enough to explain the experimentally measured effect.

It should be noted that these XY-MR estimates are sensitive to the values of conical domain dimensions used and to the accuracy of the $g_{MR}$ value determined by Lippmann and Kuhrt [11]; nevertheless, our analysis shows clearly that the measured XY-MR can be fully explained as being geometric in origin. Moreover, the carrier mobilities thereby inferred from the XY-MR measurements are of the same order of magnitude as those found through Z-MR measurements.

**Section S10: Domain wall current**

The conduction AFM images in figure 1 in the main text show significant smearing of the current across the nanodomain centre. This is a well-known resolution problem with cAFM, We show below that the current at LNO domain walls is certainly confined to the walls, as opposed to the centre of nanodomains. Figure S10 **a-d** shows dual AC resonance tracking PFM (DART-PFM) of LNO domain walls. These domains are poled using an AFM tip top electrode, with various pulse duration (increasing along the x-axis) and increasing pulse magnitude (along the -y-axis). S10**e,f** show conducting AFM in the same region, showing clearly that the enhanced conduction is confined to the domain walls.

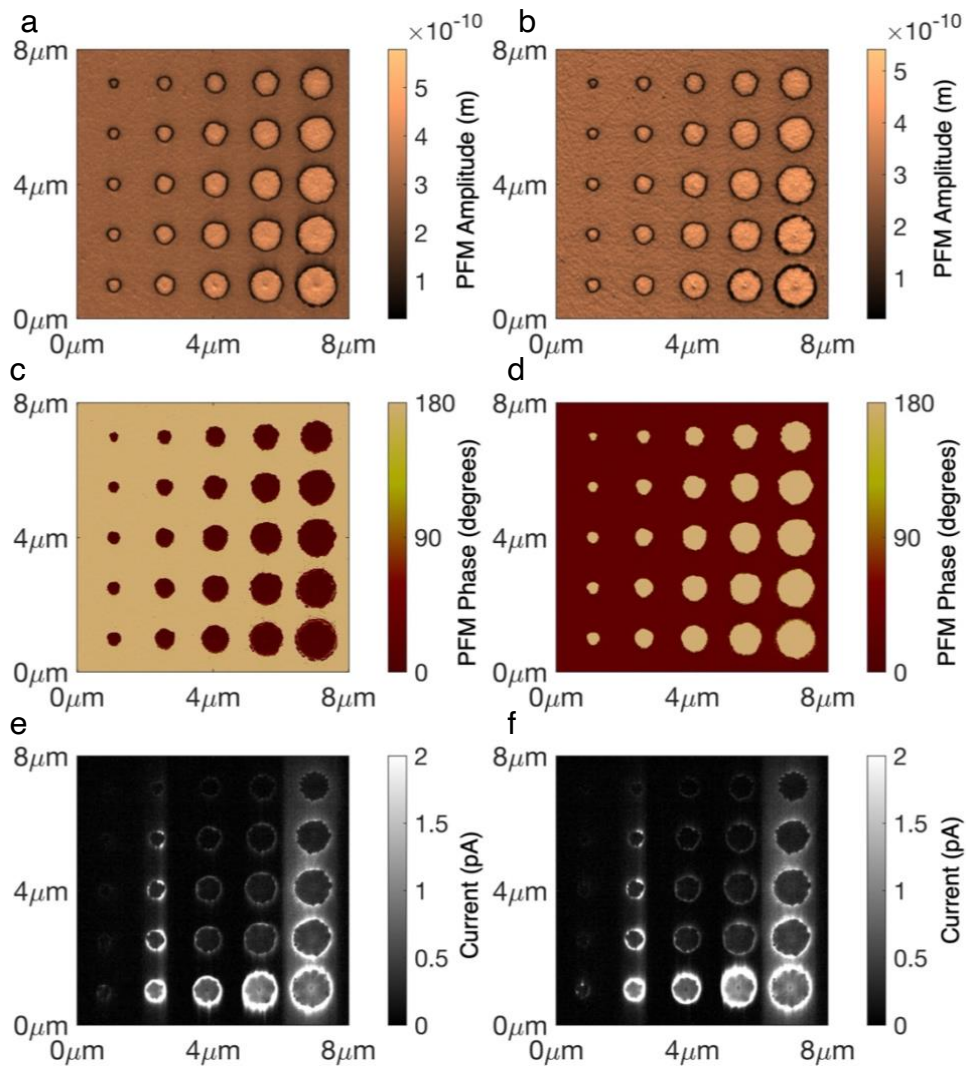

**Figure S10 | Domain wall currents**. **a, b**, PFM amplitude and **c, d**, PFM phase, taken in dual resonance tracking PFM (DART-PFM) mode. These domains are poled using an AFM tip top electrode, with various pulse duration (increasing along the x-axis) and increasing pulse magnitude (along the -y-axis). **e, f** Conduction-AFM from the same regions as **a-d**.